\documentclass[twoside,11pt]{article}

\usepackage{blindtext}

\usepackage[abbrvbib, preprint]{jmlr2e}

\usepackage{jmlr2e}
\usepackage[utf8]{inputenc} 
\usepackage[T1]{fontenc}    
\usepackage{hyperref}       
\usepackage{url}            
\usepackage{booktabs}       
\usepackage{amsfonts}       
\usepackage{amsmath}
\usepackage{amssymb}
\usepackage{nicefrac}       
\usepackage{microtype}      
\usepackage[dvipsnames]{xcolor}

\usepackage{graphicx}

\usepackage{lastpage}
\jmlrheading{24}{2023}{1-\pageref{LastPage}}{2/23; Revised
7/23}{12/23}{23-0158}{Paula Harder, Alex Hernandez-Garcia, Venkatesh Ramesh, Qidong Yang, Prasanna Sattegeri, Daniela Szwarcman, Campbell Watson, David Rolnick}
\ShortHeadings{title}{Harder, Hernandez-Garcia, Ramesh, Yang, Sattegeri, Szwarcman, Watson, Rolnick}

\ShortHeadings{Hard-Constrained Climate Downscaling}{Harder et al.}
\firstpageno{1}

\begin{document}

\title{Hard-Constrained Deep Learning for Climate Downscaling}

\author{\name Paula Harder \email paula.harder@mila.quebec \\
       \addr Fraunhofer ITWM, Kaiserslautern, Germany\\
       Mila Quebec AI Institute, Montreal, Canada
       \AND
       \name Alex Hernandez-Garcia\thanks{Equal contribution.}\\
       \addr Mila Quebec AI Institute, Montreal, Canada\\
       \addr University of Montreal, Montreal, Canada
       \AND
       \name Venkatesh Ramesh\footnotemark[1]\\
       \addr Mila Quebec AI Institute, Montreal, Canada\\
       University of Montreal, Montreal, Canada
       \AND
       \name Qidong Yang\\
       \addr Mila Quebec AI Institute, Montreal, Canada\\
       \addr New York University, New York, USA
       \AND
       \name Prasanna Sattegeri\\
       \addr IBM Research,  New York, USA
       \AND
       \name Daniela Szwarcman\\
       \addr IBM Research, Brazil
       \AND
       \name Campbell D. Watson\\
       \addr IBM Research, New York, USA
       \AND
       \name David Rolnick\\
       \addr Mila Quebec AI Institute, Montreal, Canada\\
       McGill University, Montreal, Canada
       }

\editor{Shakir Mohamed}

\maketitle

\begin{abstract}
The availability of reliable, high-resolution climate and weather data is important to inform long-term decisions on climate adaptation and mitigation and to guide rapid responses to extreme events. Forecasting models are limited by computational costs and, therefore, often generate coarse-resolution predictions. Statistical downscaling, including super-resolution methods from deep learning, can provide an efficient method of upsampling low-resolution data. However, despite achieving visually compelling results in some cases, such models frequently violate conservation laws when predicting physical variables. In order to conserve physical quantities, here we introduce methods that guarantee statistical constraints are satisfied by a deep learning downscaling model, while also improving their performance according to traditional metrics. We compare different constraining approaches and demonstrate their applicability across different neural architectures as well as a variety of climate and weather data sets. Besides enabling faster and more accurate climate predictions through downscaling, we also show that our novel methodologies can improve super-resolution for satellite data and natural images data sets.
\end{abstract}

\section{Introduction}
Accurate modeling of weather and climate is critical for taking effective action to combat climate change. In addition to shaping global understanding of climate change, local and regional predictions guide adaptation decisions and provide the impetus for action to reduce greenhouse gas emissions \citep{TheOngoingNeed}. Predicted and observed quantities such as precipitation, wind speed, and temperature impact decisions in sectors such as agriculture, energy, and transportation. While these quantities are often required at a fine geographical and temporal scale to ensure informed decision-making, most climate and weather models are extremely computationally expensive to run (sometimes taking months even on super-computers), resulting in coarse-resolution predictions. Thus, there is a need for fast methods that can generate high-resolution data based on the low-resolution models that are commonly available.

The terms \textit{downscaling} in climate science and \textit{super-resolution} (SR) in machine learning (ML) refer to a map from low-resolution (LR) input data to high-resolution (HR) versions of that same data; the high-resolution output is referred to as the super-resolved (SR) data. Downscaling via established statistical methods---\textit{statistical downscaling}---has been long used by the climate science community to increase the resolution of climate data \citep{maraun_widmann_2018}. In statistical downscaling, there are two subfields, \textit{perfect prognosis} and \textit{model output statistics} \citep{maraun_widmann_2018}. Whereas perfect prognosis learns the relationship between LR and HR observations, model output statistics learns directly the function from model output to observations, including a form of bias correction. 

In perfect prognosis, predictands and predictors usually include different variables. If both inputs and outputs consist of the same variables, this is referred to as super-resolution, even in a climate context. In parallel, computer vision SR has evolved rapidly using various deep learning architectures, with such methods now including super-resolution convolutional neural networks (CNNs) \citep{srcnn}, generative adversarial models (GANs) \citep{esrgan}, vision transformers \citep{sr_transformer}, and normalizing flows \citep{srflow}. Increasing the temporal resolution via frame interpolation is also an active area of research for video enhancement \citep{deep_voxel} that can be transferred to spatiotemporal climate data. Recently, deep learning approaches have been applied to a variety of climate and weather data sets, covering both model output data and observations. In addition to using neural networks to learn parametrization, replace model parts in a hybrid setup, or run full forecasts, downscaling is a field for deep learning to improve and accelerate Earth system simulations \citep{reichstein2019}. Climate super-resolution has mostly focused on CNNs \citep{vandal2017}, recently shifting toward GANs \citep{stengel_2020, wang_2021}.

Most statistical downscaling tools are applied offline as a tool for post-processing. In that case, machine learning methods can be directly employed on the output data, following data reformatting. However, downscaling tools could be applied online within a global climate model too (e.g. \citet{onlineDS}), where a lower resolution output of a climate model part is downscaled, and its high-resolution version is fed back into the climate model. 

There are certain tasks that are more suited for hard-constraining than others. One important point is that there exists a relationship between low-resolution and high-resolution samples for downscaling or between input and output for other tasks, given by an equation. This can be the case when modeling physical quantities, with, for example, mass or energy conservation that exists between LR and HR pairs. On the one hand, if we consider compressed or blurry images and the task is to remove the effects of compression or blur, there may be no known constraint between low and high resolution, so constraining methodologies would not be applicable. On the other hand, for some data from e.g.~satellites or telescopes, images are created by summing photons across a given field of view, so the value at a given pixel can be interpreted as the sum of values at unobserved subpixels; in such cases, hard constraints could potentially be useful.

In this work, we introduce novel methods to strictly enforce physics-inspired consistency constraints between low-resolution (input) and high-resolution (output) images. We do this via a constraint layer at the end of a neural architecture, which renormalizes the prediction either additively, multiplicatively, or with an adaptation of the softmax layer. We use climate and weather data sets based on European Center for Medium-Range Weather Forecasts (ECMWF) reanalysis data version 5 (ERA5) \citep{era5}, Weather Research and Forecast Model (WRF) data \citep{wrf}, and the Norwegian Earth System Model (NorESM) \citep{noresm} data, spanning different quantities such as water content, temperature, water vapor, and liquid water content. For the ERA5 data, we increase the resolution by different factors, we create data sets with an enhancement of factors ranging from 2 over 4 and 8 to 16. We show the utility of our methods across architectures including CNNs, GANs, CNN-RNNs, and a novel architecture that we introduce to apply super-resolution in both spatial and temporal dimensions. Besides climate data sets, we show that our methods are able to improve predictive accuracies for lunar satellite imagery super-resolution as well as on standard image super-resolution benchmark data sets, like Set5, Set14, Urban100 and BSD100. 
Our code is available at \url{https://github.com/RolnickLab/constrained-downscaling} and our main data set can be found at \url{https://drive.google.com/file/d/1IENhP1-aTYyqOkRcnmCIvxXkvUW2Qbdx/view}.

\paragraph{Contributions} Our main contributions can be summarized as follows:
\begin{itemize}
    \item We introduce a novel constraining methodology for deep learning-based downscaling methods, which guarantees that physical consistency constraints such as mass and energy conservation between low-resolution and high-resolution are satisfied.
    \item We show that our method improves predictive performance across different deep learning architectures on a variety of climate data sets.
    \item Additionally, we show that our method increases the accuracy of super-resolution in other domains, such as natural images and satellite imagery. 
    \item Finally, we introduce a new deep learning architecture for downscaling along both spatial and temporal dimensions.
\end{itemize}

\section{Related work}

\paragraph{Deep Learning for Climate Downscaling}
There exists extensive work on ML methods for climate and weather observation and prediction downscaling, from CNN architectures \citep{vandal2017} to GANs \citep{stengel_2020} and normalizing flows \citep{climalign}. 
Recently, GANs have become a very popular architecture choice, including many works on precipitation model downscaling \citep{wang_2021, watson_2020, Chaudhuri2020CLiGANAS} as well as other quantities such as wind and solar data \citep{stengel_2020}.
Unified frameworks comparing methods and benchmarks were introduced by \citet{medina_2020} to assess different SR-CNN setups and by \citet{wisosuper} with the introduction of a new data set for wind and solar SR. 
To date, there has been limited work on spatiotemporal SR with climate data. Some authors have looked at super-resolving multiple time steps at once without increasing the temporal resolution \citep{spatio_temp1,spatio_temp2}. \citet{spatio-tep3} did increase the temporal resolution by simply treating the time steps as different channels and using a standard SR-CNN.

\paragraph{Constrained Learning for Climate}
 Various works on ML for climate science have attempted to enforce certain physical constraints via soft penalties in the loss \citep{beucler2019}, linearly constrained neural networks for convection \citep{beucler}, or aerosol microphysics emulation \citep{harder2022} using completion or correction methods. \citet{zanna} and \citet{bolton} use a final fixed convolutional layer to achieve momentum and vorticity conservation in an ML ocean model. A different line of work incorporates constraints
into machine learning based on flux balances (Sturm and Wexler, 2020, 2022; Yuval et al.,
2021). These strategies use domain knowledge of how properties flow to ensure conservation
of different quantities. Instead of predicting tendencies directly, fluxes are predicted. \citet{hess} introduces one global constraint to be applied to bias-correct the precipitation prediction generated by a GAN. Outside of climate science, recent work has emerged on enforcing hard constraints on the output of neural networks (e.g.~\citet{dc3}).

\paragraph{Constrained Learning for Downscaling}
In super-resolution for turbulent flows, MeshfreeFlowNet \citep{meshfree} employs a physics-informed model which adds PDEs as regularization terms to the loss function. 
In parallel to our work, the first approaches employing hard constraints for climate-related downscaling were introduced:  \citet{geiss2023} introduced an enforcement operator applied to multiple CNN architectures for scientific data sets. A CNN with a multiplicative renormalization layer is used for atmospheric chemistry model downscaling in \citet{geiss_2022}.
We are the first to compare a variety of different hard-constraining approaches and also apply them to multiple deep learning architectures.

\section{Enforcing constraints}

When modeling physical quantities such as precipitation or water mass, principled relationships such as mass conservation can naturally be established between low-resolution and high-resolution samples. Here, we introduce a new methodology to incorporate these constraints within a neural network architecture. We choose hard constraints enforced through the architecture over soft constraints that use an additional loss term. Hard constraints guarantee certain constraints even at inference time, whereas soft constraining encourages the network to output values that are close to satisfying constraints, by minimizing a penalty during training, but do not provide any guarantees. Additionally, for our case hard constraining increases the predictive ability, and soft constraining can lead to unstable training and an accuracy-constraints trade-off \citep{harder2022}. Adding hard constraints restricts the hypothesis space to a smaller subspace that satisfies the constraints. With that, we reformulate the learning problem to an easier problem and achieve better results including prior knowledge.

\subsection{Setup}Consider the case of downscaling low-resolution pixels $x$ by a factor of $N$ in each linear dimension, and let $n:=N^2$. Let $y_i, i = 1,\ldots , n$ be the values in the predicted high-resolution patch that correspond to $x$. The set $\{y_i\}$ for $i = 1,\ldots , n$ is also referred to as a super-pixel. Then, a conservation law takes the form of the following constraint:
\begin{equation}\label{ds_constraint}
    \frac{1}{n}\sum_{i=1}^{n}y_i = x.
\end{equation}

Depending on the predicted quantity, there may additionally be an inequality constraint associated with the data. In our work, there was only one example, concerning the positivity of several physical quantities (e.g. water mass). The inequality for this case would be:
\begin{equation}
     \forall i\in [[1,n]], y_i\geq 0.
\end{equation}

We note that the methodologies we suggest in this work only deal with this special case.


\subsection{Constraint layer}

\begin{figure}[htb]
\begin{center}
\centerline{\includegraphics[width=\columnwidth]{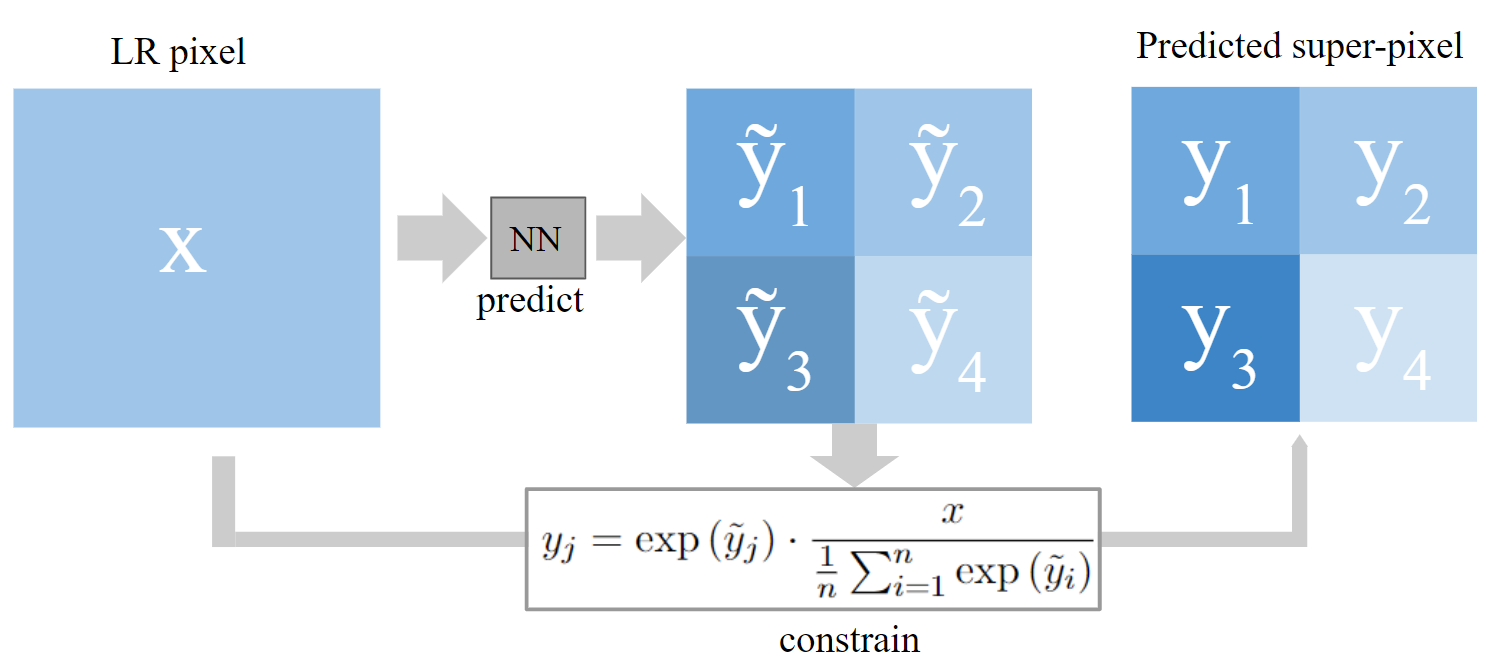}}
\caption{Our Softmax Constraining Layer (SmCL) is shown for one input pixel $x$ and the corresponding predicted $2\times 2$ super-pixel for the case of $2\times$ upsampling. This layer is added at the end of a NN and enforces given constraints guaranteed by construction. Besides equality constraints, it enforces positivity of the outputs.}
\label{pixel}
\end{center}
\end{figure}

\begin{figure}[htb]
\begin{center}
\centerline{\includegraphics[width=\columnwidth]{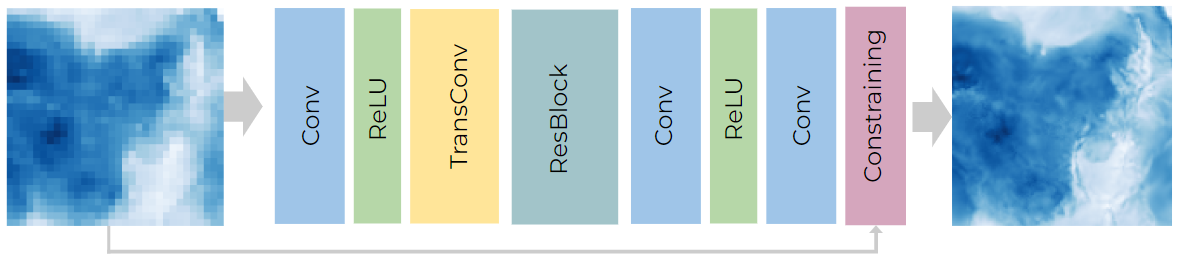}}
\caption{The CNN architecture used here for $2\times$ upsampling including the constraint layer (in red). The LR input is passed to the last layer, the constraint layer, to enforce the constraint and produce a consistent HR output.}
\label{cnn_arc}
\end{center}
\end{figure}

We introduce three different alternatives as constraint layers: additive constraining, multiplicative constraining, and softmax-based constraining. These are all added at the end of any neural architecture, as shown in Figure \ref{cnn_arc}, and all satisfy Eq. \ref{ds_constraint} by construction. The constraints are applied for each pair of input pixel $x$ and the corresponding SR $N\times N$ patch. An illustration is shown in Figure \ref{pixel}.
We will use $\tilde{y}_i, i = 1,\ldots , n$ to denote the intermediate outputs of the neural network before the constraint layer and ${y}_i, i = 1,\ldots , n$ to be the final outputs after applying the constraints.

\paragraph{Additive constraining}

For our Additive Constraint Layer (AddCL), we take the intermediate outputs and reset them using the following operation:
\begin{equation}\label{add_constraint}
    y_j = {\tilde{y}_j} + x-\frac{1}{n}\sum_{i=1}^{n}{\tilde{y}_i}.
\end{equation}

We also consider a more complex additive approach, the Scaled Additive Constraint Layer (ScAddCL), which was introduced in parallel work to ours by \citet{geiss2023}: 
\begin{equation}
    y_j = {\tilde{y}_j}+ (x- \frac{1}{n}\sum_{i=1}^{n}{\tilde{y}_i})\cdot \frac{\sigma+\tilde{y}_i}{\sigma+\frac{1}{n}\sum_{i=1}^{n}{\tilde{y}_i}},
\end{equation}

with $\sigma := \mbox{sign}(\frac{1}{n}\sum_{i=1}^{n}{\tilde{y}_i}-x)$, so $\sigma \in \{-1,1\}$ The pixel values are assumed to in $[-1,1]$. For more details see \citet{geiss2023}.

\paragraph{Multiplicative constraining}
For the Multiplicative Constraint Layer (MultCL) approach, we rescale the intermediate output using the corresponding input value $x$:
\begin{equation}\label{mult_constraint}
    y_j = {\tilde{y}_j}\cdot\frac{x}{\frac{1}{n}\sum_{i=1}^{n}{\tilde{y}_i}}.
\end{equation}

A similar approach is used in \citet{geiss_2022}. Note that this approach can violate non-negativity constraints (e.g. 18 pixels per 128x128 patch for $8\times$ upsampling, see Table \ref{cnn_results}), so it is sometimes detrimental.
Multiplicative constraining can however be generalized by introducing any function $g$:
\begin{equation}\label{multgen_constraint}
    y_j = {g(\tilde{y}_j)}\cdot\frac{x}{\frac{1}{n}\sum_{i=1}^{n}{g(\tilde{y}_i})}.
\end{equation}
If $g$ is positive, the output is guaranteed to be positive too.

\paragraph{Softmax constraining}

For predicting quantities like atmospheric water content, we want to enforce the output to be non-negative for it to be physically valid. Here, we use a softmax multiplied by the corresponding input pixel value $x$:

\begin{equation}
    y_j = \exp{(\tilde{y}_j})\cdot\frac{x}{\frac{1}{n}\sum_{i=1}^{n}\exp{(\tilde{y}_i})}.
\end{equation}

This Softmax Constraint Layer (SmCL) is a special case of Eq. (\ref{multgen_constraint}) with $g\equiv \exp$ and enforces $y_i\geq 0, i = 1,\ldots , n$. 

\paragraph{Differences of Constraint Layers}
The four different constraint layers have in common that they all enforce Eq. (\ref{ds_constraint}) by construction and we will see in Section \ref{sec:results} that the differences in performance are rather small. To point out and summarize the differences: Whereas ScAddCl ($[-1,1]$) and MultCL (non-zero) are restricted in the range of input values they can handle, AddCL and SmCL work with any inputs. SmCL gives only positive outputs, which can be either beneficial by serving as an additional physical constraint or too restrictive if the output domain includes negative values. MultCL might get unstable for values close to zero. Additionally, the choice of constraint layer influences the variance among super-pixels, with SmCL having the highest variance (see Table \ref{variance}):

\subsection{Generalization of our constraining methodologies}

The focus of this work is on a consistency constraint for downscaling, but the methodology is not limited to this and can be applied to different setups. It can be slightly adapted to e.g. enforce a weighted formulation of Eq. (1), global constraint, or mass conservation constraints for emulation. Here we show how our constraint layers can be employed for different cases, starting with a more general setup and then formulating special relevant cases.

\subsubsection{Generalization setup}
We consider the learning task (supervised or unsupervised), where $X\in \mathbb{R}^{n_{\text{in}}}$ is our input and $y\in \mathbb{R}^{n_{\text{out}}}$ the final output. Let $(I_j)_{j=1,\ldots ,n_{p}}$ be a partition of $\{1,\ldots,n_{\text{out}}\}$ into $n_{p}$ subsets ($n_p$ determines how many different constraints are imposed, e.g. $n_{\text{in}}$ for our downscaling setup), $g_{ij}:\mathcal{D}\subset \mathbb{R}\rightarrow \mathbb{R}, \ i \in I_j$ an invertible function and $h_j:\mathbb{R}^{n_{\text{out}}}\rightarrow\mathbb{R} $ an arbitrary function. The set of constraints is given by
\begin{equation}\label{gen_constraint}
    \sum_{i \in I_j} g_{ij}(y_i) = h_j(X),
\end{equation}
for each $j=1,\ldots,n_p$.

These constraints can then be enforced with the above-introduced layers restated as follows
\begin{eqnarray*}
    y_i^{\text{AddCL}} & = & g_{ij}^{-1}({\tilde{y}_i} + \frac{1}{n}h_j(X)-\frac{1}{n}\sum_{k\in I_j}{\tilde{y}_k}), \\
    y_i^{\text{MultCL}} & = & g_{ij}^{-1}( {\tilde{y}_i}\cdot\frac{h_j(X)}{\sum_{k\in I_j}{\tilde{y}_k}}),\\
    y_i^{\text{SmCL}} & = & g_{ij}^{-1}( \exp{(\tilde{y}_i)}\cdot\frac{h_j(X)}{\sum_{k\in I_j}{\exp{(\tilde{y}_k})}}),
\end{eqnarray*}
for $i \in I_j$ and $j =1,\ldots,n_p$.

The main case considered in this work (Eq. (\ref{ds_constraint}) is a special case with $h_j(X) = nX_j$ for $j$ indexing all super-pixels and $g$ being the identity function. Note that MultCl and SmCL cannot be directly applied if $h_j\equiv 0$ for any $j$, leading to a constant prediction.

\subsubsection{Weighted formulation}

In an Earth system modeling context data often originates from a latitude-longitude grid. This implies that the areas in each field are not exactly the same. The downscaling consistency constraint (Eq. (\ref{ds_constraint})) is then changed to a weighted formulation:

\begin{equation}\label{weighted_ds_constraint}
    \frac{1}{n}\sum_{i=1}^{n}\alpha_i y_i = x.
\end{equation}

Analogously, the AddCL, MultCl, and SmCL are reformulated as
\begin{eqnarray*}\label{wadd_constraint}
    y_i^{\text{AddCL}} & = & \frac{1}{\alpha_i}({\tilde{y}_i} + x-\frac{1}{n}\sum_{i=k}^{n}{\tilde{y}_k}).\\
    y_i^{\text{MultCL}} & = & {\tilde{y}_i}\cdot\frac{x}{\alpha_i \frac{1}{n}\sum_{k=1}^{n}{\tilde{y}_k}}\\
    y_i^{\text{SmCL}} & = & {\exp{(\tilde{y}_i})}\cdot\frac{x}{\alpha_i \frac{1}{n}\sum_{k=1}^{n}\exp{(\tilde{y}_k})}
\end{eqnarray*}

We note that in our case we do not use a weighted formulation, since the ERA5 LR data is created by average pooling without weighting and the WRF data covers a small area, so there the lat-lon cells have about the same area.

\subsubsection{Relaxing constraints and global constraining}
The constraint layers can be relaxed by increasing the constraint window size; this can then impose soft constraints. In the extreme case, this would reduce the number of constraints to one and gives the possibility of adding global constraint. The constraints would be the same as in Eq. (\ref{ds_constraint}), but with $n$ being the number of total pixels.

\subsubsection{Application in emulation}
Our constraining methodology is not limited to downscaling and can enforce mass conservation e.g. in emulation tasks. An example could be aerosol microphysics emulation \citep{harder2022}, where different aerosol masses need to be conserved within each time step. The predicted aerosol masses among different size bins $y_i,i\in I_\text{dust}$ for a specific aerosol type, eg. dust, have to add up to the sum of the input aerosol masses $X_i, i \in I_{dust}$ of the same species:
\begin{equation*}\label{loss}
    \sum_{i\in I_{\text{dust}}} y_i = \sum_{i\in I_{\text{dust}}} X_i
\end{equation*}
This conservation of mass can be enforced with the AddCL, MultCl, or SmCl:
\begin{eqnarray*}
    y_i^{\text{AddCL}} & = & {\tilde{y}_i} + \sum_{k\in I_{\text{dust}}} X_k-\sum_{k\in I_{\text{\text{dust}}}}{\tilde{y}_k} \\
    y_i^{\text{MultCL}} & = &  {\tilde{y}_i}\cdot\frac{ \sum_{k\in I_{\text{dust}}} X_k}{\sum_{k\in I_{\text{dust}}}{\tilde{y}_k}}\\
    y_i^{\text{SmCL}} & = & \exp{(\tilde{y}_i)}\cdot\frac{ \sum_{k\in I_{\text{dust}}} X_k}{\sum_{k\in I_{\text{dust}}}{\exp{(\tilde{y}_k})}}
\end{eqnarray*}
Here, SmCL again would additionally guarantee positive masses.

\section{Data}

To test and evaluate our proposed method, we create a variety of data sets as well as use existing and established ones. We generate multiple data sets based on the ERA5 data using average pooling to create the LR inputs, which has been the standard methodology in climate downscaling studies (see e.g.~\citet{spatio-tep3,spatio_temp2}). We also use data sets based on the outputs of models such as the Weather and Research Forecasting (WRF) Model and the Norwegian Earth System Model (NorESM) that contain real low-resolution simulation data matched to high-resolution data. Finally, we test our methods on non-climate data sets: lunar satellite imagery and natural images. An overview of all the different data sets used can be found in Table \ref{data}.

\begin{table}[htb] 
\caption{The different data sets we use to test our constraint layers. The names are given to identify the data sets throughout the paper. Most data sets are based on ERA5 atmospheric water content data and LR is generated synthetically, we include different upsampling factors, an ood case, and temporal data sets. Additional data sets include the moist static energy (MEn) data set as well as WRF and NorESM model data. Lunar and natural images give non-climate application data sets. The results for data sets in bold can be found in the main paper the rest is given in the appendix for improved focus and clarity.}
\label{data}
\vskip 0.15in
\begin{center}
\begin{small}
\begin{sc}
\begin{tabular}{lcccc}
\toprule
Name & Source & Type & Dim. & Size  \\
& &  & LR/HR & train/val/test  \\
\midrule
TCW2 & ERA5 & water cont. &  (1,64,64)/(1,128,128) & 40k/10k/10k \\
\bf{TCW4} & ERA5 & water cont. &  (1,32,32)/(1,128,128) & 40k/10k/10k \\
TCW8 & ERA5 & water cont. &  (1,16,16)/(1,128,128) & 40k/10k/10k \\
TCW16 & ERA5 & water cont. &  (1,8,8)/(1,128,128) & 40k/10k/10k \\
TCW OOD & ERA5 & water cont. &  (1,32,32)/(1,128,128) & 40k/10k/10k \\
\bf{TCW T1} & ERA5 & water cont. &  (3,32,32)/(3,128,128) & 40k/10k/10k \\
\bf{TCW T2} & ERA5 & water cont. &  (2,32,32)/(3,128,128) & 40k/10k/10k \\
MEn & ERA5 & water vapor &  (3,32,32)/(3,128,128) & 40k/10k/10k \\
 & & liq. water & & \\
  & & temp. & & \\
\bf{WRF} & WRF & temp. &  (1,45,45)/(1,135,135) & 20k/4k/4k \\
NorESM & NorESM & temp. &  (1,32,32)/(1,64,64) & 24k/12k/12k \\
Lunar & satell. & photons &  (1,32,32)/(1,128,128) & 132k/16k/16k \\
Nat & Nat. images &  RGB &(3,128,128)/(3,512,512) & var. \\
\bottomrule
\end{tabular}
\end{sc}
\end{small}
\end{center}
\vskip -0.1in
\end{table}

\subsection{ERA5 data set}
The ERA5 data set \citep{era5} is a so-called \textit{reanalysis} product from the ECMWF that combines model data with worldwide observations. The optimal physical model state that best fits the observations is found through the process of data assimilation. ERA5 is available as global, hourly data with a $0.25^{\circ}\times0.25^{\circ}$ resolution, which is roughly $25~\mbox{km}$ per pixel in the mid-latitudes. It covers all years starting from 1950. 

\begin{figure}[htb]
\begin{center}
\centerline{\includegraphics[width=\columnwidth]{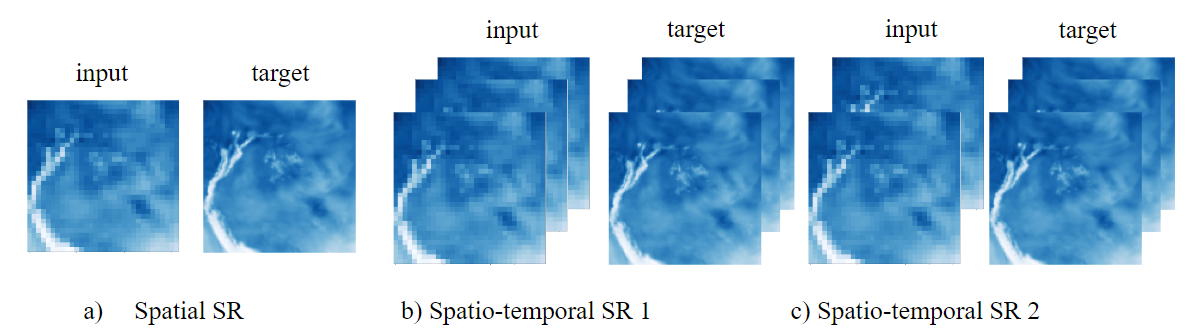}}
\caption{Samples of the three different data set types used in this work. a) A data pair we use for our standard spatial super-resolution task. The input is an LR image and the target is the HR version of that. b) A data pair for performing SR for multiple time steps simultaneously. The input is a time series of LR images and the output is the same time series in HR. c) A data pair where SR is performed both temporally and spatially, with two LR time steps as input and 3 HR time steps as a target.}
\label{data sets}
\end{center}
\end{figure}

\paragraph{Total water content data set}

For this work, the quantity we focus on is the total column water (tcw) that is given in $\text{kg}/\text{m}^2$ and describes the vertical integral of the total amount of atmospheric water content, including water vapour, cloud water, and cloud ice but not precipitation.

\paragraph{Spatial SR data}
To obtain our high-resolution data points we extract a random $128\times128$ pixel image from each available time step (each time step is $721\times1440$ and there are roughly 60,000 time steps available). We randomly sample 40,000 data points for training and 10,000 for each validation and testing. The low-resolution counterparts are created by taking the mean over $N\times N$ patches, where $N$ is our upsampling factor. A sample pair is shown in Figure \ref{data sets} a). This operation is physically sound, considering that conservation of water content means that the water content (density per squared meter) described in an LR pixel should be equal to the average of the corresponding HR pixels. We can also observe in LR-modeled data such as WRF data (see below) that the modeled quantities in a low-resolution run are approximately the mean of a high-resolution run, which further justifies our coarsening strategy. 

\paragraph{Spatio-Temporal data sets}
Including the temporal evolution of our data, we create two additional data sets. For the first data set, one sample consists of 3 successive time steps, the same time steps for both input and target, but at different resolutions. This is done to perform spatial SR for multiple time steps simultaneously, see Figure  \ref{data sets} b). We select three random $128\times128$ pixel areas per global image, resulting in the same number of examples as the procedure described above. We split the data randomly as before, and each time step is downsampled by taking the spatial mean. 
We then create a second data set, that is built for the learning task of increasing both spatial and temporal dimensions. We again crop three images out of a series of three successive time steps to obtain our high-resolution target. To create the low-resolution input, we decrease both temporal and spatial dimensions. To decrease the temporal resolution, we remove the intermediate (the second) time step in each sample, i.e. perform sub-sampling. To decrease the spatial resolution we apply the same operation as before, i.e. compute the mean spatially. These results result in two LR inputs, see Figure  \ref{data sets} c). Temporally coarse-graining by subsampling not by averaging is done to avoid leakage of future information into previous time steps

\paragraph{OOD data set}
For the data sets described above, the train-val-test split is done randomly. To understand how our constraining influences out-of-distribution generalization, we create a data set with a split in time. Here, we expect patterns to appear in the later time steps that are out-of-distribution of what was previously observed. We train on older data and then test on more recent years: for training, we use the years 1950-2000, for validation 2001-2010, and for final testing 2011-2020.

\paragraph{Energy data set}
Also originating from the ERA5 data, we create a second data set including different physical variables coming with different constraints as well.  This data set is constructed to preserve moist static energy and water masses while predicting water vapor, liquid water content, and air temperature. The variables are taken from the pressure level at 850hPa.

\subsection{WRF data}

In \citet{watson_2020}, a data set using the Advanced Research version of the WRF Model is introduced. It comprises hourly operational weather forecast data for Lake George in New York, USA from 2017-01-01 to 2020-03-20. More details about the model and its configuration can be found in \citet{watson_2020}. The variable we consider for this work is the temperature at 2m above the ground. Unlike the previous data sets, this one does not involve synthetic downsampling but includes two forecasts run at different resolutions with different physics-based parameterizations: one at 9 km horizontal resolution and one at 3 km. Our goal is to predict the 3 km resolution temperature field given the 9 km one and builds on work by \citet{wrf}, which used the same data set.

\subsection{Constraints in our data sets}

In predicting distinct physical quantities, there are different constraints we need to consider. Most of our data sets include the downscaling constraints given by (\ref{ds_constraint}), which are satisfied by the LR-HR pairs either approximately (for simulations that are run at LR and HR with quantities respecting physical conservation laws) or exactly (in the case of average pooling for creating the LR version). We detail the constraints in the following subsections.

\paragraph{Water content conservation}
For predicting the total column-integrated water content, we are given the low-resolution water content $Q^{(LR)}$ and must obtain the super-resolved version $Q^{(SR)}$. The downscaling constraint or mass conservation constraint (\ref{ds_constraint}) for each LR pixel $ q^{(LR)}$ and the corresponding super-pixel $(q_i^{(SR)})_{i=1,\ldots ,n}$ is then given by
\begin{equation}\label{eq:mass_con}
    \frac{1}{n}\sum_{i=1}^{n}q_i^{(SR)} = q^{(LR)}.
\end{equation}

\paragraph{Moist static energy conservation}
One of our tasks includes predicting column-integrated water vapor, liquid water, and temperature while conserving both water mass and moist static energy. As described above, water mass conservation is straightforward, directly applying our constraining methodology. On the other hand, the (column-integrated) moist static energy $S$ is approximated by:
\begin{equation}\label{eq:moist_satic}
    S \approx ((1-Q_v)\cdot c_{pd} + Q_L\cdot c_l)\cdot T + L_v\cdot Q_v,
\end{equation}

where $$ L_v \approx 2.5008 \cdot 10^6 + (c_{pw}-c_L)\cdot(T-273.16)$$ is the latent heat of vaporization in $ (Jkg^{-1})$. The water vapor $Q_v [kg\cdot kg^{-1}]$, the liquid water $Q_L [kg\cdot kg^{-1}]$, and the temperature $T [K]$ are being predicted, whereas $c_{pd}, c_{pv}$ and $c_L [J\cdot K^{-1} \cdot kg^{-1}]$ are heat capacity constants. 

We use the following procedure to predict these quantities while conserving moist static energy:
\begin{enumerate}
    \item Given LR $T^{LR}, Q_V^{LR}, Q_L^{LR}$
    \item Calculate LR $S^{LR}$ with (\ref{eq:moist_satic})
    \item Predict SR $S^{SR}, Q_v^{SR}, Q_L^{SR} $ while enforcing (\ref{ds_constraint}) using one of our constraint layers
    \item Calculate SR $T^{SR}$ using (\ref{eq:moist_satic}) and SR $S^{SR}, Q_v^{SR}, Q_L^{SR} $.
\end{enumerate}
This means we predict $T^{SR}$ not directly, but by predicting $S^{SR}$. We are then able to predict the temperature $T$ while ensuring (approximate) energy conservation by applying our constraint layer to the prediction of $S^{SR}$.

\begin{figure}[htb]
\begin{center}
\vskip 0.1in
\centerline{\includegraphics[width=\columnwidth]{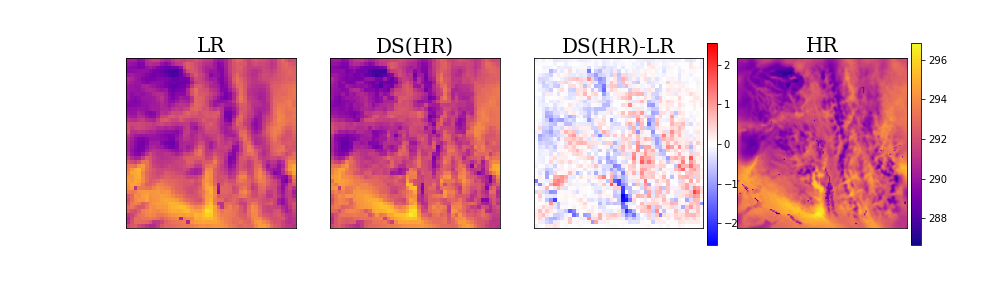}}
\caption{A LR-HR pair from the WRF temperature data. HR and LR come from different runs using the same model at different resolutions. Here we compare the real LR with the low-resolution data created by average pooling of the HR, written as DS(HR). It shows that there is not an exact match between LR and downsampled HR, which makes the success of a constraint layer more difficult. The violation of the downscaling constraint in the WRF data set is 0.684 on average.}
\label{wrf_ds}
\end{center}
\vskip -0.2in
\end{figure}

\paragraph{Different simulations}
If the LR-HR pairs are not created by taking the local mean of the HR but by using two simulations run at different resolutions, the downscaling constraint is not automatically satisfied in the data. This is the case for our WRF and NorESM data sets (NorESM data is discussed in the appendix; here, we focus on WRF). Even though the downscaling constraint is not exactly obeyed (see Figure \ref{wrf_ds}), it is approximately, and we can still apply our constraining in the same way as before. If the real low-resolution data and the downsampled high-resolution data are not significantly dissimilar, constraining can still benefit the predictive ability.

\section{Experimental setup}

We conduct two sets of experiments:
\begin{enumerate}
    \item Show the applicability of our constraining method to different neural network architectures.
    \item Show the applicability of our constraining method to different data sets and different constraint types.
\end{enumerate}
   
In most of our experiments, we use synthetic low-resolution data created by applying average pooling to the original high-res samples, as is usually done to test perfect prognosis downscaling setups. Additionally, we consider cases with pairs of real low-res and high-res simulations to show that our methods work in the intended final application. 

\subsection{Architectures}

We test our constraint methods throughout a variety of standard deep learning SR architectures including an SR CNN, conditional GAN, a combination of an RNN and CNN for spatio-temporal SR, and a new architecture combining optical flow with CNNs/RNNs to increase the resolution of the temporal dimension. The original, unconstrained versions of these architectures then also serves as a comparison for our constraining methodologies.

\paragraph{SR-CNNs}
Our SR CNN network, similar to \citet{esrcnn}, consists of convolutional layers using $3\times 3$ kernels and ReLU activations. The upsampling is performed by a transpose convolution followed by residual blocks (convolution, ReLU, convolution, adding the input, ReLU). The architecture for $2\times$ downscaling is shown in Figure \ref{cnn_arc}.

\paragraph{SR-GAN}
A conditional GAN architecture \citep{cgan} is a common choice for super-resolution \citep{srgan}. Our version uses the above-introduced CNN architecture as the generator network. The discriminator is used from \citep{srgan}, it consists of convolutional layers with a stride of 2 to decrease the dimensionality in each step, with ReLU activation. It is trained as a classifier to distinguish SR images from real HR images using a binary cross-entropy loss. The generator takes as input both Gaussian noise as well as the LR data and then generates an SR output. It is trained with a combination of an MSE loss, helping reconstruction, and the adversarial loss given by the discriminator, like a standard SR GAN, e.g.~\citet{ledig2017photo}.

\paragraph{SR-ConvGRU}
We apply an SR architecture based on the GAN presented by \citet{spatio_temp2}, which uses ConvGRU layers to address the spatio-temporal nature of super-resolving a time series of climate data. Here, we use the generator on its own, both during inference and training time without the discriminator, providing a deterministic approach.

\paragraph{SR-FlowConvGRU}

\begin{figure}[htb]
\begin{center}
\centerline{\includegraphics[width=\columnwidth]{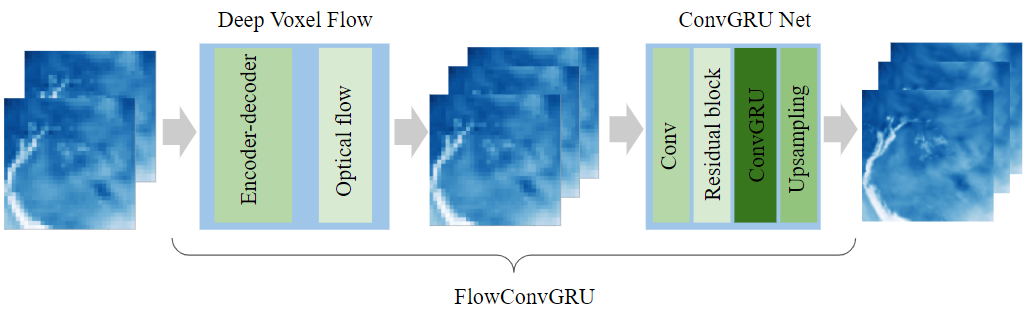}}
\caption{Our novel spatio-temporal architecture, combining Deep Voxel Flow and a ConvGRU. The inputs are two LR images at two times, the first part predicts the in-between time step using the Deep Voxel Flow model, the second part increases the spatial resolution of the three time steps using a Convolutional GRU net.}
\label{voxel_arc}
\end{center}
\end{figure} 
To increase the temporal resolution of our data we employ the Deep Flow method \citep{deep_voxel}, a deep learning architecture for video frame interpolation combining optical flow methods with neural networks. We introduce a new architecture combining the Deep Flow model and the ConvGRU network (FlowConvGRU): First, we increase the temporal resolution resulting in a higher-frequency time-series of LR images on which we then apply the ConvGRU architecture to increase the spatial resolution. The combined neural networks are then trained end-to-end. The architecture is shown in Figure \ref{voxel_arc}.

\subsection{Training}

Our models were trained with the Adam optimizer, a learning rate of $0.001$, and a batch size of 256. We trained for 200 epochs, which took about 3---6 hours on a single NVIDIA A100 Tensor Core GPU, depending on the architecture. All models use the MSE as their criterion, the GAN additionally uses its discriminator loss term. All the data are normalized between 0 and 1 for training, except for the cases where the ScAddCL is applied. In the case of this constraint layer we scale the data between -1 and 1 as proposed in \citet{geiss2023}. For our time-dependent models though, ConvGRU and FlowConvGRU, we are scaling between 0 and 1, because the original scaling led to NaN-values during training.

\subsection{Baselines}

\paragraph{Pixel enlargement}
This baseline consists of scaling the LR input to the same size as the HR by duplicating the pixels. We include this to have reference metrics that reflect how close the LR is to the HR data. This baseline conserves mass by construction.

\paragraph{Bicubic upsampling}
As a simple non-ML baseline, we use bicubic interpolation for spatial SR and take the mean of two frames for temporal SR.

\paragraph{Soft constraining}
Soft-constraining has been successfully applied before to a variety of physics-informed deep-learning tasks. Here we use it to see how it compares to hard constraints. Soft-constraining is done by adding a regularization term to the loss function. Our MSE loss is then changed to the following:

\begin{equation}
   \text{Loss} = (1-\alpha) \cdot \text{MSE} + \alpha\cdot \text{Constraint violation},
\end{equation}

\noindent where the constraint violation is the mean overall constraint violations between an input pixel $x$ and the corresponding super-pixel $y_i, i=1,\ldots ,n$:
\begin{equation}
\text{Constraint violation} = \text{MSE}\left(\frac{1}{n}\sum_{i=1}^{n}y_i,\; x\right) .
\end{equation}
We conducted an experiment to investigate the impact of $\alpha$ values on final model performance; the results are reported in the appendix. For our main paper we choose $\alpha = 0.99$.

\paragraph{Unconstrained counterparts}
Furthermore, we always compare against an unconstrained version of the above-introduced standard SR NN architectures (SR-CNN, SR-GAN, SR-ConvGRU, SR-FlowConvGRU).

\paragraph{Clipping}
We also run the standard CNN, but with clipping applied at inference. This is a common practice to remove negative values. Results can be found in the appendix, see Table \ref{clip_results}. This method does not guarantee mass conservation nor significantly improves performance.

\section{Results and discussion}
\label{sec:results}

\begin{figure}[htb]
\begin{center}
\centerline{\includegraphics[width=\textwidth]{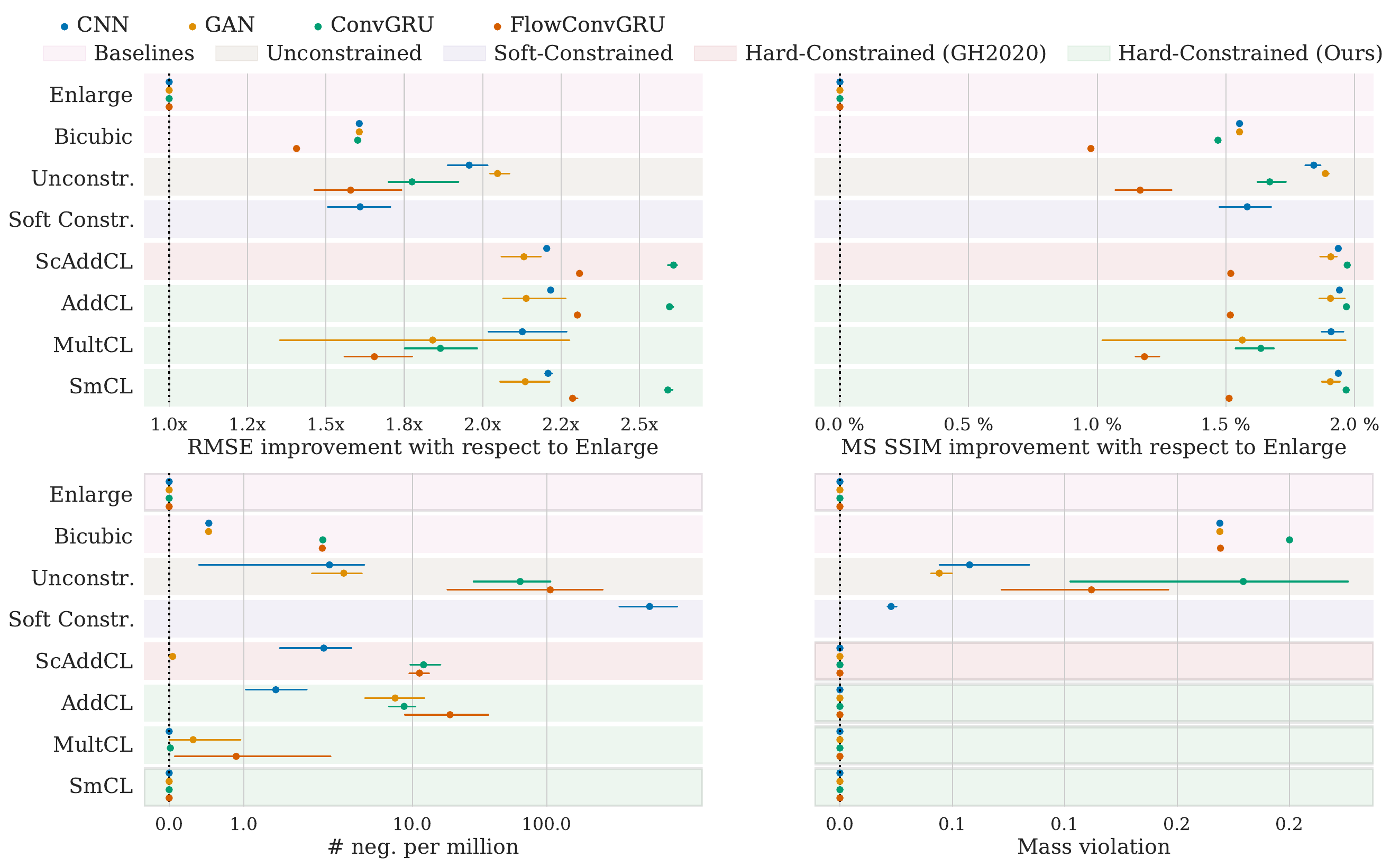}}
\caption{Metrics for different constraining methods and architectures applied to the water content data sets (TCW4, TCW T1 and TCW T2), calculated over 10,000 test samples. The mean and confidence interval from 3 runs are shown, for RMSE and MS-SSIM relative to the Enlarge baseline for number of negative pixels (per mil.) and mass conservation violation the absolute values are shown. The framed box indicates that the method achieves zero violation of the physics, no negative pixels or mass conservation up to numerical precision. Tables with more metrics can be found in the appendix}
\label{fig:main}
\end{center}
\end{figure}

For evaluating our results, we use typical metrics for weather and climate super-resolution: root-mean-square error (RMSE), mean absolute error (MAE) and mean bias as well as typical metrics for super-resolution: peak signal-to-noise ratio (PSNR), structural similarity index measure (SSIM), multi-scale SSIM (MS-SSIM), Pearson correlation and Fractional Skill Score (FSS). We show RMSE and MS-SSIM in the main paper, while the others can be found in the appendix. Most metrics are highly correlated in our case. For the GAN giving a probabilistic prediction, we also use continuous ranked probability score (CRPS). Because we are interested in the violation of conservation laws and predicting non-physical values, we also look at the average constraint violation, the number of (unwanted) negative pixels, and the average magnitude of negative values. We additionally look at the variance among the pixels within a predicted super-pixel and investigate the difference for constraining methods. The key results are aggregated in Figure \ref{fig:main}.

\subsection{Different constraining methods}
Whereas hard-constraining shows exact conservation and appears to enhance performance, the application of soft-constraining on the other hand does decrease constraint violation, but still maintains a significant magnitude of it, which can be seen in Figure \ref{fig:main} for example. Also, soft-constraining seems to suffer from an accuracy-constraints trade-off, where depending on the regularization factor $\alpha$, either the constraint violation is reduced, or the accuracy increases, but it struggles to do both simultaneously. A table for different $\alpha$ is shown in the appendix. 
Among the hard-constraining methodologies, the multiplicative renormalization layer, MultCL, performs the weakest in terms of predictive skills (see Figure \ref{fig:main}), which could be due to instability when inputs get close to zero. The three other methods, ScAddCL, AddCL, and SmCL, often have very similar measurements. SmCL shows the advantage of also enforcing positivity when necessary (see Figure \ref{fig:main}). ScAddCL divides the number of violation by more than 2 compare to the AddCL and MulCL gets close to zero violation in many cases.

\subsection{Different architectures}

As shown in Figure \ref{fig:main} for all architectures (CNN, GAN, ConvGRU, FlowConvGRU), adding the constraint layers enforces the constraint and improves the evaluation metrics compared to the CNN case. Constraining the GAN leads to less of a performance boost, but AddCL and SmCL still enhance the predictions compared to the unconstrained GAN. Including the temporal dimensions, the constraining improves the prediction quality much more significantly than in the case with just a single time step (see Figure \ref{fig:main}). 

\subsection{Different data sets and constraints}
The success of our constraining methodology does not depend on the upsampling factor: in Table \ref{cnn_results}, we can see that the constraining methods work well and improve all metrics for upsampling factors of 2, 4, 8, and 16.
When applied to our out-of-distribution data set, the improvement achieved by adding constraints is even more pronounced than for the randomly split data (see results in the appendix). The constraints can help architectures with their generalization ability.


Not only mass can be conserved, but other quantities such as moist static energy. We show that moving on to different quantities of the ERA5 data set, temperature, water vapor, and liquid water. Looking at Table \ref{energy_results} (see appendix), one can observe similar results for liquid water $Q_L$ and water vapor $Q_v$ as for the total water content: ScAddCL, AddCL, and SmCL significantly improve results in all measures over the unconstrained CNN, while enforcing energy and mass conservation. For temperature, on the other hand, MultCL performs the strongest, followed by SmCL, whereas AddCL and ScAddCL achieve smaller improvements in the scores.

%

Our WRF temperature data set includes low-resolution data points drawn from a separate simulation, rather than downsampling, and therefore it results in much harder tasks. Table \ref{wrf_results} shows that the scores are improved slightly with our constraint layer, this might be counterintuitive given there is a violation in the training data, but this violation is relatively small, it appears like random noise, so no bias is introduced. This way the constraints again lead to a simpler learning problem and are able to improve performance. The fact, that the constraints are slightly violated in the original data set could motivate soft-constraining, but nevertheless, we can observe that soft-constraining harms the predictive performance, while hard-constraining is surprisingly beneficial. The constraint violation in the original data has an RMSE of 0.6838 on average.
\begin{table}[htb] 
\caption{We show four metrics for different constraining methods applied to the SR CNN applied on the WRF temperature data, calculated over 10,000 test samples. We choose the most common (RMSE, MAE, SSIM) and relevant (constr. viol) for our cases. The mean is taken over 3 runs. The best scores are highlighted in bold blue.}
\label{wrf_results}
\vskip 0.15in
\begin{center}
\begin{small}
\begin{sc}
\begin{tabular}{lllcccc}
\toprule
Data & Model & Constraint & RMSE & MAE & MS-SSIM &  Constr. viol. \\
\midrule
WRF & Enlarge & none & 1.015 & 0.648 & 94.51 & \bf{\color{blue}0.000} \\
WRF &CNN & none & 0.952& 0.618& 94.92&0.181 \\
WRF &CNN & soft & 1.020& 0.660& 94.57&0.032 \\
WRF &CNN & SmCL & \bf{\color{blue}0.950} & \bf{\color{blue} 0.592} & \bf{\color{blue}95.25} & \bf{\color{blue}0.000} \\
\bottomrule
\end{tabular}
\end{sc}
\end{small}
\end{center}
\vskip -0.1in
\end{table}

\begin{figure*}[htb]
\begin{center}
\vskip 0.1in
\centerline{\includegraphics[width=\columnwidth]{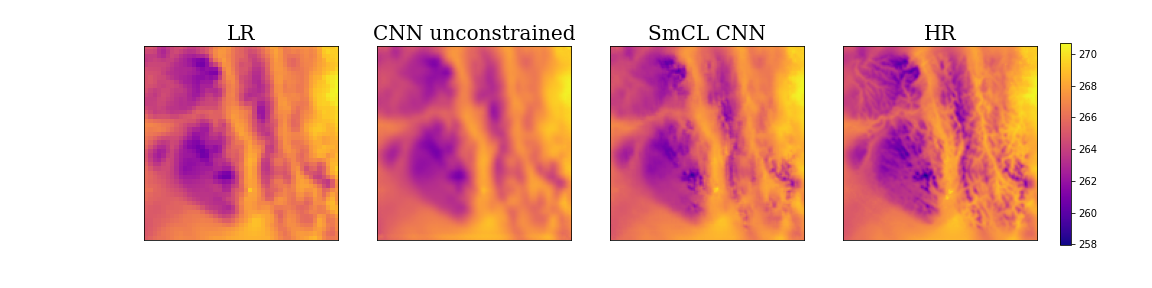}}
\caption{A random prediction for the WRF temperature test data set. We compare unconstrained and softmax-constrained predictions. It can be seen that in this case, the constraining improves the visual quality significantly including more fine-grain details.}
\label{wrf}
\end{center}
\vskip -0.2in
\end{figure*} 

Finally, we also show that applying our constraint methodology can improve results in other domains, even in cases where there is no physics involved. We see that both for the lunar satellite imagery and the natural images benchmark data sets, the application of our SmCL improves the traditional metrics, as shown in Tables \ref{lunar_results} and \ref{sr_results}. 

\subsection{Perceptual quality of predictions}

\begin{figure*}[htb]
\begin{center}
\vskip 0.1in
\centerline{\includegraphics[width=\columnwidth]{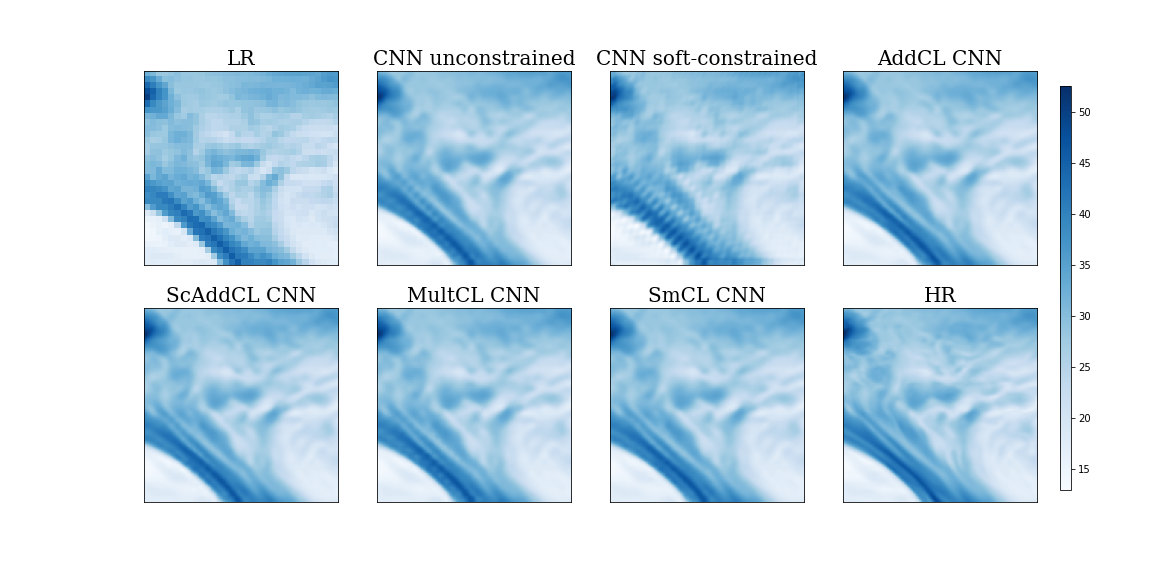}}
\caption{One example image from the test set. Shown here are the LR input, different constrained and unconstrained predictions, and the HR image as a reference. This example is from the TCW4 test data set. For the unconstrained CNN prediction, we can observe some artifacts in the lower left part, which get amplified by applying soft-constraining but decreased using hard-constraining like AddCL, ScAddCL, or SmCl.}
\label{plots}
\end{center}
\vskip -0.2in
\end{figure*} 

\begin{figure}[htb]
\begin{center}
\vskip 0.1in
\centerline{\includegraphics[width=\columnwidth]{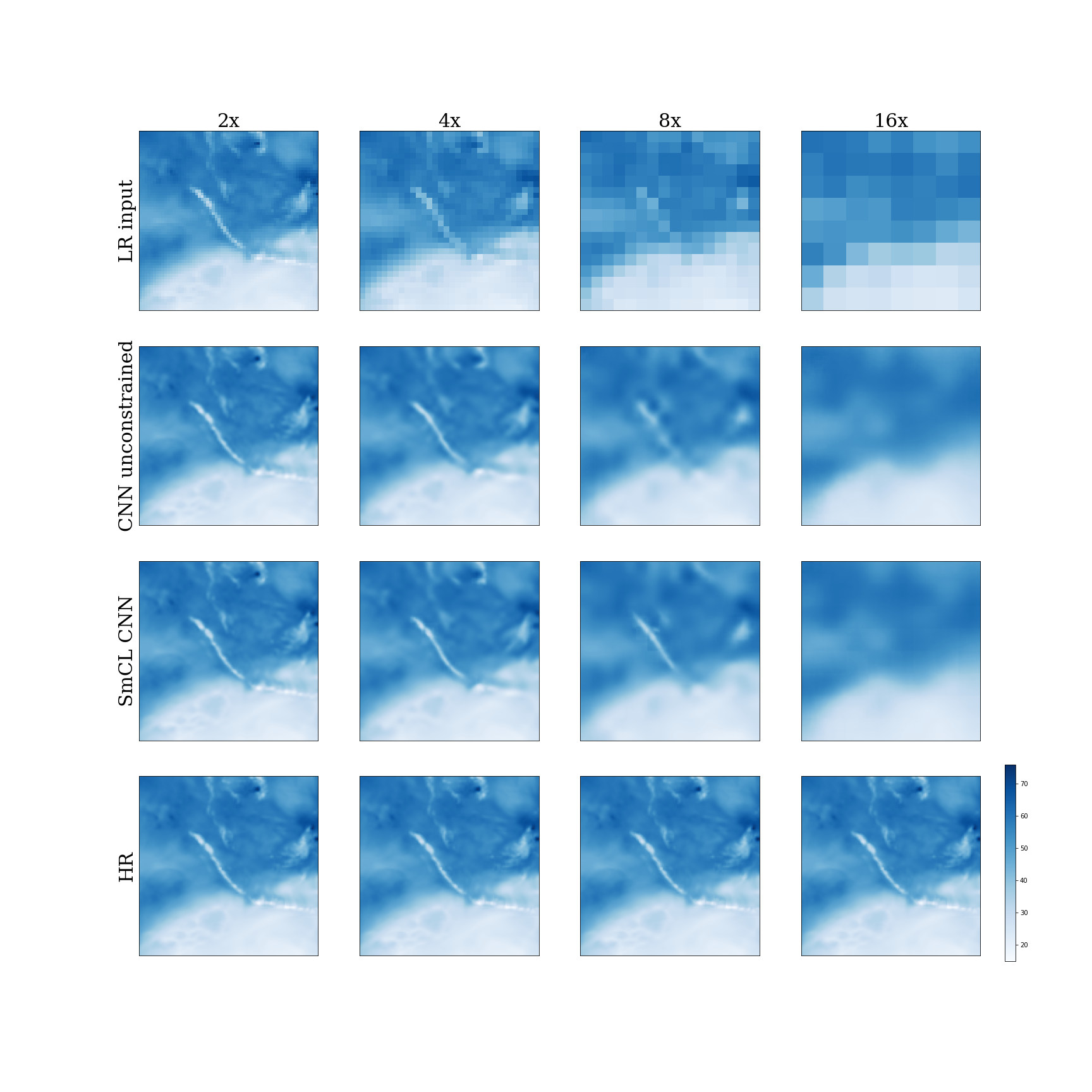}}
\caption{One example image is chosen randomly from the test set. Each model was trained for the same target resolution but with a different upsampling factor. The first row shows the LR inputs for each resolution and the last row the corresponding HR ground truth. The second and third rows show the prediction of an unconstrained CNN and with the SmCL, respectively.}
\label{factors}
\end{center}
\vskip -0.2in
\end{figure}

\begin{figure*}[htb]
\begin{center}
\vskip 0.1in
\centerline{\includegraphics[width=\columnwidth]{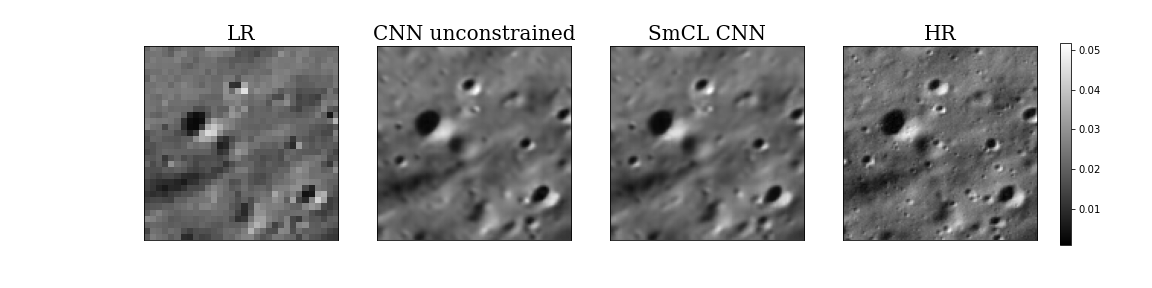}}
\caption{A random sample prediction from the lunar data set is shown. We compare the unconstrained with the constrained prediction.}
\label{lunar}
\end{center}
\vskip -0.2in
\end{figure*}

Additionally to an enhancement quantitatively, we can see an improved visual quality for some examples, as shown in Figure \ref{plots} and \ref{factors} for the water content data. For the WRF temperature forecast data, we see a very significant improvement in the perceptual quality of the prediction. Looking at an example, such as shown in Figure \ref{wrf}, we can see how much more detail is added to the prediction when adding our constraining. For the lunar satellite imagery, Figure \ref{lunar} shows that applying constraints can make the image slightly less blurry.

\subsection{Development of error during training}
\begin{figure*}[htb]
\begin{center}
\vskip 0.1in
\centerline{\includegraphics[width=\columnwidth]{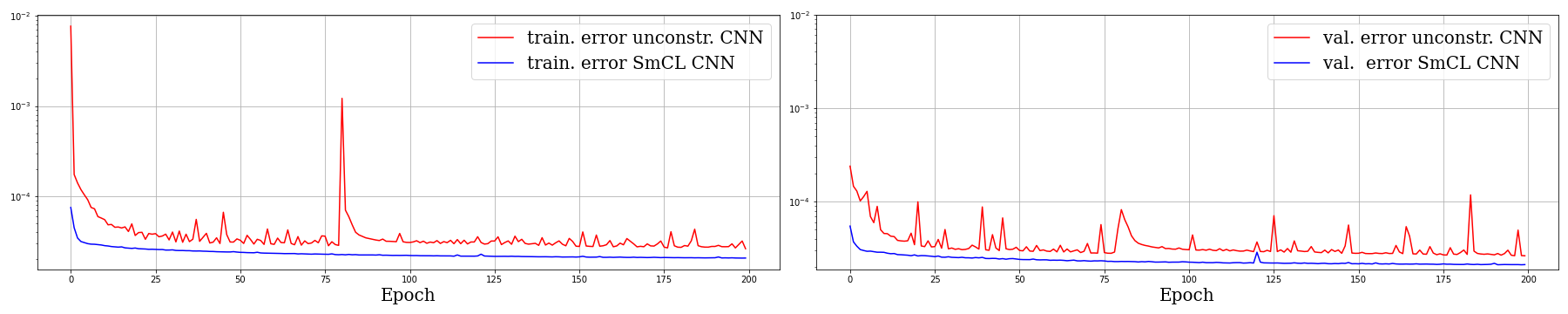}}
\caption{The development of training and validation errors with increasing iterations during training. Shown for an unconstrained CNN and CNN+SmCL applied to the water content data. We can observe how hard constraining accelerates convergence and smooths the learning curve, both measured in training and validation error.}
\label{curve}
\end{center}
\vskip -0.2in
\end{figure*}

Observing how the MSE develops during training (see Figure \ref{curve}), we can see that the curve of the constrained network is generally lower than the unconstrained one. Additionally, it can be seen that constraining helps smooth both the training and validation curves.

\subsection{Spatial distribution of errors}
\begin{figure*}[htb]
\begin{center}
\vskip 0.1in
\centerline{\includegraphics[width=\columnwidth]{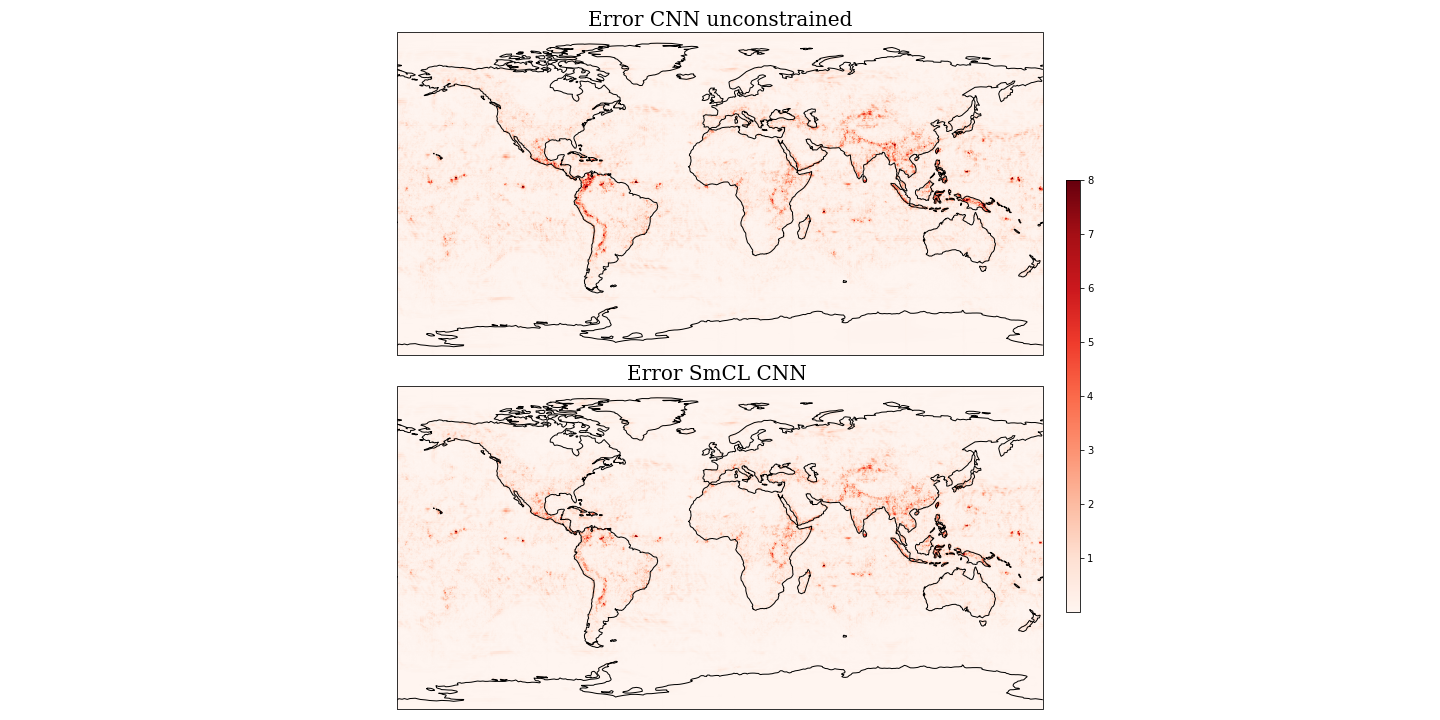}}
\caption{The errors of the global predictions for unconstrained and constrained (SmCL) CNNs, when compared to the ground truth. The CNN is applied per $32\times32$ patch and then put together for a global predictions at a random time step. Used here is the TCW4 data set. We can observe how the stronger errors in coastal and mountainous regions for the unconstrained predictions are dampened by soft-max constraining.}
\label{map_error}
\end{center}
\vskip -0.2in
\end{figure*}

\begin{figure*}[htb]
\begin{center}
\vskip 0.1in
\centerline{\includegraphics[width=\columnwidth]{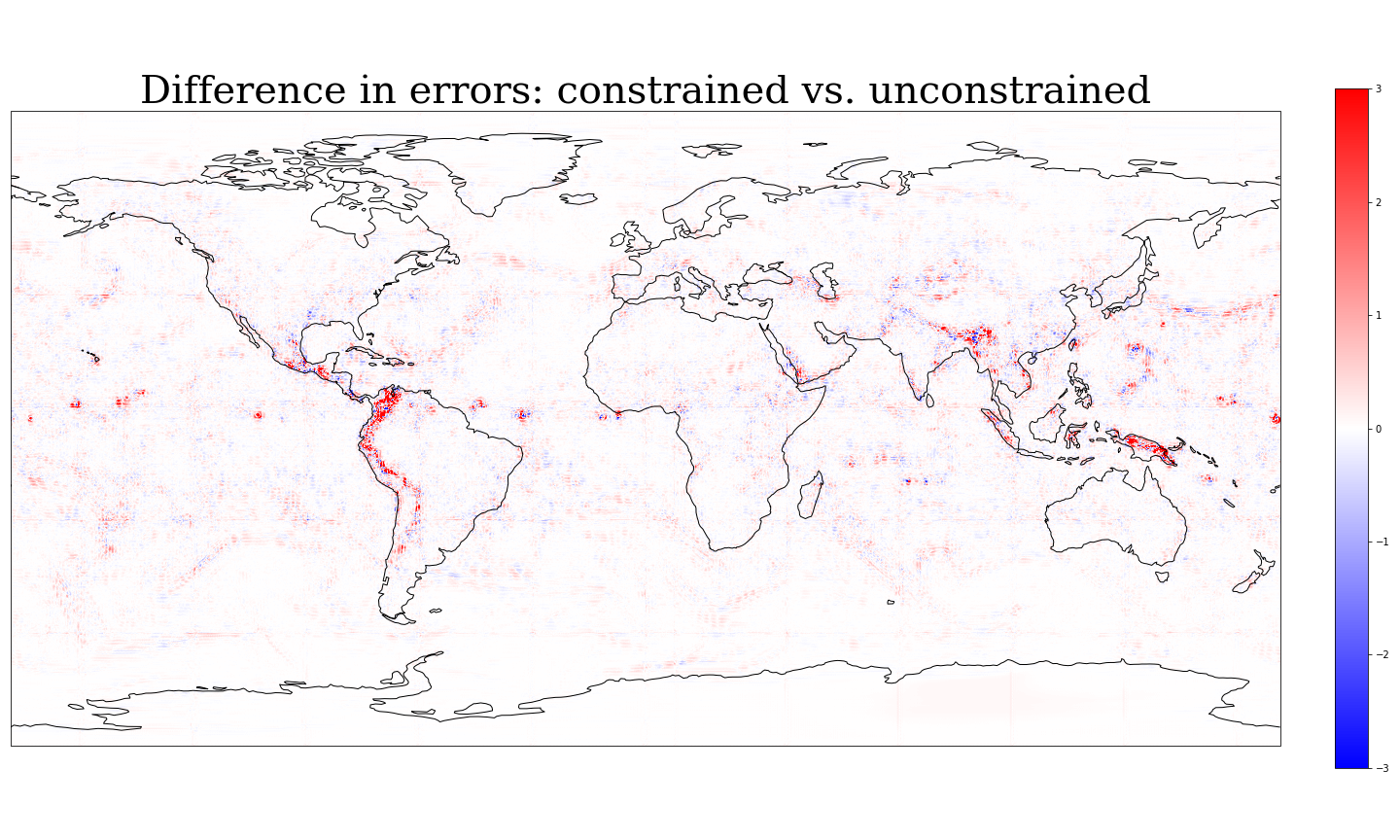}}
\caption{The difference in the errors of constrained and unconstrained predictions from Figure \ref{map_error}. Positive values (red) mean a higher error in the unconstrained version. We trim values at 3, so everything that has a difference greater than 3 is shown as full red for better visibility.}
\label{map_error_error}
\end{center}
\vskip -0.2in
\end{figure*}

A known issue in downscaling methods is the so-called coastal effect, where errors of predictions tend to be more pronounced in coastal regions. Besides coastal region areas, mountain ridges can also be critical. In Figure \ref{map_error}, we show the error of the unconstrained prediction for water content and the softmax-constrained prediction. We can see that both predictions show more errors in coastal and mountainous regions. However, if we analyze the difference in errors between the unconstrained and constrained versions, we can see in Figure \ref{map_error_error} that constraining leads to lower errors in those areas.

\subsection{Limitations}
In the case of our WRF data set, we have seen that the constraining methodology can improve predictive performance even if the underlying constraints are slightly violated by the original data. In cases where low-resolution and its high-resolution counterpart are too far apart, our model is not always able to increase the predictive skill. We built a data set from two different resolutions of the Norwegian Earth System Model (NorESM) \citep{noresm}, and applying our constraining methods improved the visual similarity of the predictions, but decreased the predictive ability. We provide scores and plots in the appendix. In the case of other sampling strategies such as subsampling spatially, our methods are not applicable in their current form and they depend on having constraints that can be formulated with Eq. (\ref{gen_constraint}).


\section{Conclusion and future work}

This work presents a novel methodology to incorporate physics-inspired downscaling hard constraints into neural network architectures for climate super-resolution. We show that this method performs well across different deep learning architectures, upsampling factors, predicted quantities, and data sets. We demonstrate its effectiveness both on standard downscaling data sets and on data created by independent simulations. Our constrained models are not only guaranteed to satisfy consistency such as mass conservation between LR and HR, but also increase predictive performance across metrics and use cases. Compared to soft-constraining through the loss function, our methodology does not suffer from the common accuracy-constraints enforcement trade-off. Our hard-constraining performance enhancement is not only limited to climate super-resolution but also noticeable in satellite imagery of the lunar surface as well as standard benchmark data sets of natural images. Within the climate context, our constraint layer can help with common issues connected to deep learning applied to downscaling: it dampens the coastal effect, errors get lower in critical regions, out-of-distribution generalization is improved and training can be more stable. Hard-constraining can weaken performance if the enforced relationships are strongly violated in the true data (see NorESM data). If a bias exists in the LR (or other input) it can be propagated to the HR prediction by constraining on the LR.

Future work could extend the application of our constraint layer to other climate-related tasks beyond downscaling. Climate model emulation (e.g. \citet{beucler} and \citet{harder2021emulating}) for example could strongly benefit from a reliable and performance-enhancing method to enforce physical laws. For post-processing purposes, the offline application of our method, our code is readily available. To deploy these constrained super-resolution methods online, the next step is to use Fortran-Python bridges (e.g. \citep{ott2020fortran}) to include them in global climate model runs. 

\color{black}

\vskip 0.2in

\section*{Acknowledgement and Disclosure of Funding}
PH acknowledges the funding received by the Fraunhofer Institute for Industrial Mathematics. DR was funded in part by the Canada
CIFAR AI Chairs Program. The authors also are grateful for support from the NSERC Discovery Grants program, material support from NVIDIA in the form of computational resources, and technical support from
the Mila IDT team in maintaining the Mila Compute Cluster.

\newpage

\section*{Appendix A: Tuning soft-constraining}

Here we investigate the influence of the factor $\alpha$ on the soft-constraining method in more detail. Table \ref{alpha} shows how the increase of $\alpha$ improves the mass conservation but only up to a value between $0.014$ and $0.017$. At the same time, it shows that the predictive skill decreases with the increase of $\alpha$ significantly.
\begin{table*}[htb] 
\caption{Metrics  calculated over 10,000 validation samples. The best scores are highlighted in bold blue, second best in bold black.}
\label{alpha}
\vskip 0.15in
\begin{center}
\begin{small}
\begin{sc}
\begin{tabular}{llccccc}
\toprule
Data & Alpha & RMSE & MAE & MS-SSIM &  Mass viol. & \#Neg\\
 & & & & & & per mil.\\
\midrule
TCW4 & 0.0001 & \bf{0.241}& \bf{0.102}& 99.95&0.021 & 1.21\\
TCW4 & 0.001 & \bf{\color{blue}0.237}& \bf{\color{blue}0.100}& \bf{\color{blue}99.96}& 0.022 & \bf{\color{blue}0.12}\\
TCW4 &  0.01 & 0.247 & 0.103 & \bf{\color{blue}99.96} & 0.022 & 1.39 \\ 
TCW4 &   0.1 & 0.252 & 0.104 & 99.95 & 0.023 & \bf{0.41}\\
TCW4 & 0.9  & 0.268 & 0.110 & 99.95 & 0.020 & 16.83\\
TCW4 &  0.99 & 0.297& 0.133 & 99.94& \bf{\color{blue}0.014} & 31.01\\
TCW4 &  0.999 & 0.477 & 0.261 & 99.84 &\bf{0.016} & 600.96 \\
TCW4 &  0.9999 & 0.706 & 0.433 & 99.71 & 0.017 & 3867.90 \\
TCW4 &  1 & 2.618 & 1.814 & 94.22 & 0.017 & 960.42\\
\bottomrule
\end{tabular}
\end{sc}
\end{small}
\end{center}
\vskip -0.1in
\end{table*}

\section*{Appendix B: Clipping for nonnegativity}

As natural RGB images have a well-defined range, it is common in CNN and GAN implementations to clip the pixels at inference time to the desired range, removing negative values, for example. Here, in Table \ref{clip_results} we show that doing that gives a very small increase in performance, but still performs significantly worse than SmCL, which achieves also zero negative values. We want to point out that a combination of a constraint layer such as MultCL and clipping would lead to the clipping layer to destroy the enforced consistency given by the contraint layer if applied afterwards.

\begin{table}[htb] 
\caption{Metrics for different constraining methods applied to the SR CNN + clipping applied on the water content data set, calculated over 10,000 test samples. The mean is taken over 3 runs. The best scores are highlighted in bold blue.}
\label{clip_results}
\vskip 0.15in
\begin{center}
\begin{small}
\begin{sc}
\begin{tabular}{lllccccc}
\toprule
Data & Model & Constraint & RMSE & MAE & MS-SSIM &  Mass viol. & \#Neg\\
 & & & & & & per mil.\\
\midrule
TCW4 &CNN & none & 0.661 & 0.327 &99.39&0.059 & 2.41 \\
TCW4 &CNN & clip & 0.657 & {0.326} & {99.440} & {0.058} & \bf{\color{blue}0}\\
TCW4 &CNN & SmCL & \bf{\color{blue}0.582} & \bf{\color{blue}0.291} & \bf{\color{blue}99.49} & \bf{\color{blue}0.000} & \bf{\color{blue}0}\\
\bottomrule
\end{tabular}
\end{sc}
\end{small}
\end{center}
\vskip -0.1in
\end{table}

\section*{Appendix C: Score tables}
We show the tables with the mean scores that are displayed as Figures in the main paper and additionally include the MAE.

\


\begin{table}[htb] 
\caption{Metrics for different constraining methods applied to an SR CNN, calculated over 10,000 test samples of the water content data. The mean is taken over 3 runs. The best scores are highlighted in bold blue, second best in bold.}
\label{cnn_results}
\vskip 0.15in
\begin{center}
\begin{small}
\begin{sc}
\begin{tabular}{llllccccc}
\toprule
Data & Fact. & Model & Constraint & RMSE & MAE & MS-SSIM &  Mass viol. & \#Neg \\
 & & & & & & per mil.\\
\midrule
TCW2 & 2x & Enlarge & none & 0.422 & 0.361 & 99.61 & \bf{\color{blue}0.000} & \bf{\color{blue}0}\\
TCW2 &  2x & Bicubic & none & 0.322 & 0.137 & 99.90 & 0.066 & 0.25 \\
TCW2 & 2x & CNN & none & 0.251& 0.105& 99.95&0.026 & 1.40\\
TCW2 & 2x & CNN & soft & 0.301 & 0.137 & 99.23 & 0.016 & 104.65\\ 
TCW2 &  2x &CNN & AddCL  & 0.216 & \bf{0.092} & \bf{\color{blue}99.96} & \bf{\color{blue}0.000} & 1.31\\
TCW2 &  2x &CNN & ScAddCL & \bf{\color{blue}0.199} & \bf{\color{blue}0.0876} & \bf{\color{blue}99.96} & \bf{\color{blue}0.000} & 0.02\\
TCW2 &  2x &CNN & MultCL & 0.223 & 0.094 & \bf{\color{blue}99.96}&  \bf{\color{blue}0.000} & \bf{\color{blue}0}\\
TCW2 &  2x &CNN & SmCL & \bf{0.215} & 0.094 & \bf{\color{blue}99.96} & \bf{\color{blue}0.000} & \bf{\color{blue}0} \\
\midrule
TCW4 & 4x & Enlarge & none & 1.286 & 0.717 & 97.60&  \bf{\color{blue}0.000}& \bf{\color{blue}0}  \\
TCW4 &  4x & Bicubic & none & 0.800 & 0.401 & 99.12 & 0.169 & 0.53  \\
TCW4 &  4x & CNN & none & 0.657 & 0.326 & 99.40 & 0.058 & 2.41\\
TCW4 &  4x & CNN & soft & 0.801 & 0.410 & 99.15 & 0.023 & 581.54\\
TCW4 &  4x &CNN & AddCL & \bf{0.580} & \bf{0.290} & \bf{\color{blue}99.50} &  \bf{\color{blue}0.000} & 1.42\\
TCW4 & 4x &CNN & ScAddCL & \bf{\color{blue}0.575} & \bf{\color{blue}0.289} & \bf{\color{blue}99.50} & \bf{\color{blue}0.000} & 0.07\\
TCW4 &  4x &CNN & MultCL & 0.606 & 0.300 & 99.47 & \bf{\color{blue}0.000} & \bf{\color{blue}0}\\
TCW4 &  4x &CNN & SmCL & 0.582 & 0.291 & 99.49 & \bf{\color{blue}0.000}& \bf{\color{blue}0} \\
 \midrule
TCW8 &  8x & Enlarge & none &2.181 & 1.294 & 92.39 & \bf{\color{blue}0.000} & \bf{\color{blue}0}\\
TCW8 &  8x & Bicubic & none & 1.557 & 0.900& 96.49 & 0.318 & 6.56 \\
TCW8 &  8x & CNN & none & 1.358 & 0.782 & 97.15&0.109 & 15.48\\
TCW8 &  8x &CNN & soft & 1.640 & 0.965 & 96.06 & 0.029 & 103,702 \\
TCW8 &  8x &CNN & AddCL & \bf{1.267} & \bf{\color{blue}0.733} & \bf{\color{blue}97.41} &  \bf{\color{blue}0.000} & 632.32\\
TCW8 &  8x &CNN & ScAddCL &  \bf{\color{blue}1.264} & 0.734 & \bf{\color{blue}97.41} &  \bf{\color{blue}0.000} & 0.15\\
TCW8 &  8x &CNN & MultCL & 1.331 & \bf{\color{blue}0.733} & 97.22 & \bf{\color{blue}0.000}  &0.10\\
TCW8 &  8x &CNN & SmCL & 1.268 & 0.734 & 97.40 & \bf{\color{blue}0.000} & \bf{\color{blue}0}\\
  \midrule
TCW16 &   16x & Enlarge & none & 3.425 & 2.159 & 85.55 &   \bf{\color{blue}0.000} & \bf{\color{blue}0}\\
TCW16 &  16x & Bicubic & none & 2.723 & 1.730 & 91.72 & 0.510 & 53.67 \\
TCW16 &  16x & CNN & none & 2.450 & 1.545 & 92.68 & 0.203 & 4.15\\
TCW16 &  16x & CNN & soft & 2.794 & 1.776 & 90.74 & 0.036 & 2250.77\\
TCW16 &  16x &CNN & AddCL & \bf{\color{blue}2.364} & \bf{\color{blue}1.491} & \bf{\color{blue}92.96} & \bf{\color{blue}0.000} & 457.34\\
TCW16 & 16x &CNN & ScAddCL & \bf{2.368} & {1.495} & {92.94} & \bf{\color{blue}0.000} & 2.12\\
TCW16 &  16x &CNN & MultCL & 2.409 & 1.518 & 92.77 & \bf{\color{blue}0.000} & 0.17\\
TCW16 & 16x &CNN & SmCL & \bf{2.368} & \bf{1.492} & \bf{92.95} & \bf{\color{blue}0.000} & \bf{\color{blue}0}\\
\bottomrule
\end{tabular}
\end{sc}
\end{small}
\end{center}
\vskip -0.1in
\end{table}

\begin{table}[htb] 
\caption{Metrics for different constraining methods applied to an SR GAN, calculated over 10,000 test samples of the 4x upsampling water content data. The mean is taken over 3 runs. The best scores are highlighted in bold blue, and the second best in bold.}
\label{gan_results}
\vskip 0.15in
\begin{center}
\begin{small}
\begin{sc}
\begin{tabular}{lllcccccc}
\toprule
Data & Model & Constraint & RMSE & MAE & CRPS & MS-SSIM &  Mass viol. & \#Neg\\
 & & & & & & per mil.\\
\midrule
TCW4 & GAN & none & 0.628 & 0.313 & 0.1522 & 99.44 & 0.0453 &3.46\\
TCW4 &  GAN & AddCL  & \bf{\color{blue}0.602} & \bf{0.306} & \bf{0.1519} & \bf{\color{blue}99.46} & \bf{\color{blue}0.000} & 7.38\\
TCW4 &   GAN & ScAddCL & 0.604 & \bf{\color{blue}0.305}& \bf{\color{blue}0.1508} & \bf{\color{blue}99.46} & \bf{\color{blue}0.000} & 0.05\\
TCW4 &  GAN & MultCL & 0.732 & 0.406 & 0.1978 & 99.13 &  \bf{\color{blue}0.000} & \bf{\color{blue}0}\\
TCW4 & GAN & SmCL & \bf{0.603} & 0.310 & 0.1520 & \bf{\color{blue}99.46} & \bf{\color{blue}0.000} & \bf{\color{blue}0} \\
\bottomrule
\end{tabular}
\end{sc}
\end{small}
\end{center}
\vskip -0.1in
\end{table}

\begin{table}[htb] 
\caption{Metrics for different constraining methods applied to an SR ConvGRU, calculated over 10,000 test samples of the water content data. The best scores are highlighted in bold blue, second best in bold.}
\label{convgru_results}
\vskip 0.15in
\begin{center}
\begin{small}
\begin{sc}
\begin{tabular}{lllccccc}
\toprule
Data & Model & Constraint & RMSE & MAE & MS-SSIM &  Mass viol. & \#Neg \\
 & & & & & & per mil.\\
\midrule
TCW T1 & Enlarge & none & 1.292 & 0.718 & 97.72 & \bf{\color{blue}0.000} & \bf{\color{blue}0}\\
TCW T1& Bicubic & none &  0.807 & 0.402 & 99.16 & 0.169 & 2.16\\
TCW T1 & ConvGRU & none & 0.672 & 0.340 & 99.42& 0.102 & 55.45\\
TCW T1 & ConvGRU & AddCL  & \bf{\color{blue}0.499} & \bf{\color{blue}0.260} &  \bf{\color{blue}99.64}& \bf{\color{blue}0.000} & 1358.49\\
TCW T1 & ConvGRU & ScAddCL &  \bf{\color{blue}0.499} & \bf{\color{blue}0.260} & \bf{\color{blue}99.64} & \bf{\color{blue}0.000} & 10.58\\
TCW T1 & ConvGRU & MultCL & 0.903 & 0.472 & 98.98 &  \bf{\color{blue}0.000} & 0.25\\
TCW T1 & ConvGRU & SmCL & 0.500 & \bf{\color{blue}0.260} & \bf{\color{blue}99.64} & \bf{\color{blue}0.000} & \bf{\color{blue}0} \\
\bottomrule
\end{tabular}
\end{sc}
\end{small}
\end{center}
\vskip -0.1in
\end{table}

\begin{table}[htb] 
\caption{Metrics for different constraining methods applied to our FlowConvGRU, calculated over 10,000 test samples of the water content data set. The best scores are highlighted in bold blue, second best in bold.}
\label{Flowconvgru_results}
\vskip 0.15in
\begin{center}
\begin{small}
\begin{sc}
\begin{tabular}{lllccccc}
\toprule
Data & Model & Constraint & RMSE & MAE & MS-SSIM &  Mass viol. & \#Neg\\
 & & & & & & per mil.\\
\midrule
TCW T2 & Interpolation & none & 0.834 & 0.428 & 99.10 & 0.169 & 2.14\\
TCW T2 & FlowConvGRU & none & 0.673 & 0.352 & 99.40 & 0.072 & 18.27\\
TCW T2 & FlowConvGRU & AddCL  & \bf{\color{blue}0.509} & \bf{0.275} & \bf{\color{blue}99.63} & \bf{\color{blue}0.000} & 37.10\\
TCW T2 & FlowConvGRU & ScAddCL & \bf{\color{blue}0.509} & \bf{\color{blue}0.274} & \bf{\color{blue}99.63} & \bf{\color{blue}0.000} &13.40 \\
TCW T2 & FlowConvGRU & MultCL & 0.719 & 0.383 & 99.27 & \bf{\color{blue}0.000} & \bf{\color{blue}0}\\
TCW T2 & FlowConvGRU & SmCL & 0.514 & 0.276 & 99.62 & \bf{\color{blue}0.000} & \bf{\color{blue}0} \\
\bottomrule
\end{tabular}
\end{sc}
\end{small}
\end{center}
\vskip -0.1in
\end{table}

\begin{table}[htb] 
\caption{Metrics for different constraining methods applied to the SR CNN applied on the OOD water content data set, calculated over 10,000 test samples. The mean is taken over 3 runs. The best scores are highlighted in bold blue.}
\label{ood_results}
\vskip 0.15in
\begin{center}
\begin{small}
\begin{sc}
\begin{tabular}{lllccccc}
\toprule
Data & Model & Constraint & RMSE & MAE & MS-SSIM &  Mass viol. &\# Neg \\
 & & & & & & per mil.\\
\midrule
TCW OOD & Enlarge & none & 1.274 & 0.711 & 97.60 & \bf{\color{blue}0.000} &\bf{\color{blue}0}\\
TCW OOD &Bicubic & none & 0.792 & 0.397 & 98.63 & 0.167 & 0.55\\
TCW OOD &CNN & none & 0.661 & 0.327 &99.39&0.059 & 4.93\\
TCW OOD &CNN & AddCL & \bf{0.575} & \bf{\color{blue}0.287} &\bf{\color{blue}99.50}& \bf{\color{blue}0.000} &1.65\\
TCW OOD &CNN & ScAddCL & \bf{\color{blue}0.573} & \bf{0.288}&\bf{\color{blue}99.50}& \bf{\color{blue}0.000} &0.21\\
TCW OOD &CNN & MultCL & 0.591 & 0.294 &99.47& \bf{\color{blue}0.000} &\bf{\color{blue}0}\\
TCW OOD &CNN & SmCL & {0.579} & {0.289} & {99.49} & \bf{\color{blue}0.000} &\bf{\color{blue}0}\\
\bottomrule
\end{tabular}
\end{sc}
\end{small}
\end{center}
\vskip -0.1in
\end{table}

\begin{table}[htb] 
\caption{Metrics for different constraining methods applied to the SR CNN, calculated over the test set for water vapor, liquid water, and temperature. The mean is taken over 3 runs. For $ Q_L$, RMSE, MAE, and Constr. violation are scaled by a factor of $10^3$ for readability. The best scores are highlighted in bold blue, second best in bold.}
\label{energy_results}
\vskip 0.15in
\begin{center}
\begin{small}
\begin{sc}
\begin{tabular}{llllcccc}
\toprule
Data &Var. & Model & Constraint & RMSE & MAE & MS-SSIM &  Constr. viol. \\
\midrule
MEn &$Q_v$ & Enlarge & none & 0.474 & 0.262 & 94.74& \bf{\color{blue}0.000} \\
MEn &$Q_v$ &Bicubic & none & 0.326 & 0.182 & 97.12 & 0.07\\
MEn &$Q_v$ &CNN & none & 0.260& 0.141& 98.14&0.02 \\
MEn &$Q_v$ &CNN & AddCL  & \bf{0.250} &\bf{ 0.133} & \bf{ 98.28} & \bf{\color{blue}0.00} \\
MEn &$Q_v$ &CNN& ScAddCL & \bf{0.250} &\bf{ 0.133} &\bf{ 98.28 }& \bf{\color{blue}0.00} \\
MEn &$Q_v$ &CNN & MultCL & \bf{0.250} &\bf{ 0.133} &\bf{ 98.28} & \bf{\color{blue}0.00} \\
MEn &$Q_v$ &CNN & SmCL & \bf{\color{blue}0.248} & \bf{\color{blue}0.132} & \bf{\color{blue}98.30} & \bf{\color{blue}0.00} \\
\midrule
MEn &$Q_L$ & Enlarge & none & 0.0217 & 0.00862 & 98.34 & \bf{\color{blue}0.00000} \\
MEn &$Q_L$  &Bicubic & none & 0.0186 & 0.00765 & 98.96 & 0.00236 \\
MEn &$Q_L$  &CNN & none & 0.0157 & 0.00617& 99.15&0.00067 \\
MEn &$Q_L$  &CNN & AddCL  &\bf{\color{blue}0.0155} & \bf{\color{blue}0.00588} & \bf{\color{blue}99.18} & \bf{\color{blue}0.00000} \\
MEn &$Q_L$  &CNN & ScAddCL & \bf{\color{blue}0.0155}& \bf{\color{blue}0.00588} & \bf{99.17} & \bf{\color{blue}0.00000} \\
MEn &$Q_L$  &CNN & MultCL & 0.0166 & 0.00647 & 99.06&  \bf{\color{blue}0.00000} \\
MEn &$Q_L$  &CNN & SmCL & \bf{\color{blue}0.0155} & 0.00585 & \bf{99.17}& \bf{\color{blue}0.00000} \\
\midrule
MEn &$T$  & Enlarge & none & 0.470 & 0.288 & 99.03 & \bf{\color{blue}0.0} \\
MEn &$T$ &Bicubic & none & 0.281 & 0.156 & 99.67& 159.1\\
MEn &$T$ &CNN & none & 0.459 & 0.287 & 99.03& 139.7\\
MEn &$T$ &CNN & AddCL  & 0.276 & 0.160 & 99.67 & \bf{\color{blue}0.0} \\
MEn &$T$ &CNN& ScAddCL & 0.280 & 0.163  & 99.67  & \bf{\color{blue}0.0} \\
MEn &$T$ &CNN & MultCL & \bf{\color{blue}0.270} & \bf{\color{blue}0.155} & \bf{\color{blue}99.69} &\bf{\color{blue}0.0} \\
MEn &$T$ &CNN & SmCL &  \bf{0.272} & \bf{\color{blue}0.155} &  \bf{99.68} & \bf{\color{blue}0.0} \\
\bottomrule
\end{tabular}
\end{sc}
\end{small}
\end{center}
\vskip -0.1in
\end{table}

\begin{figure}[htb]
\begin{center}
\centerline{\includegraphics[width=\textwidth]{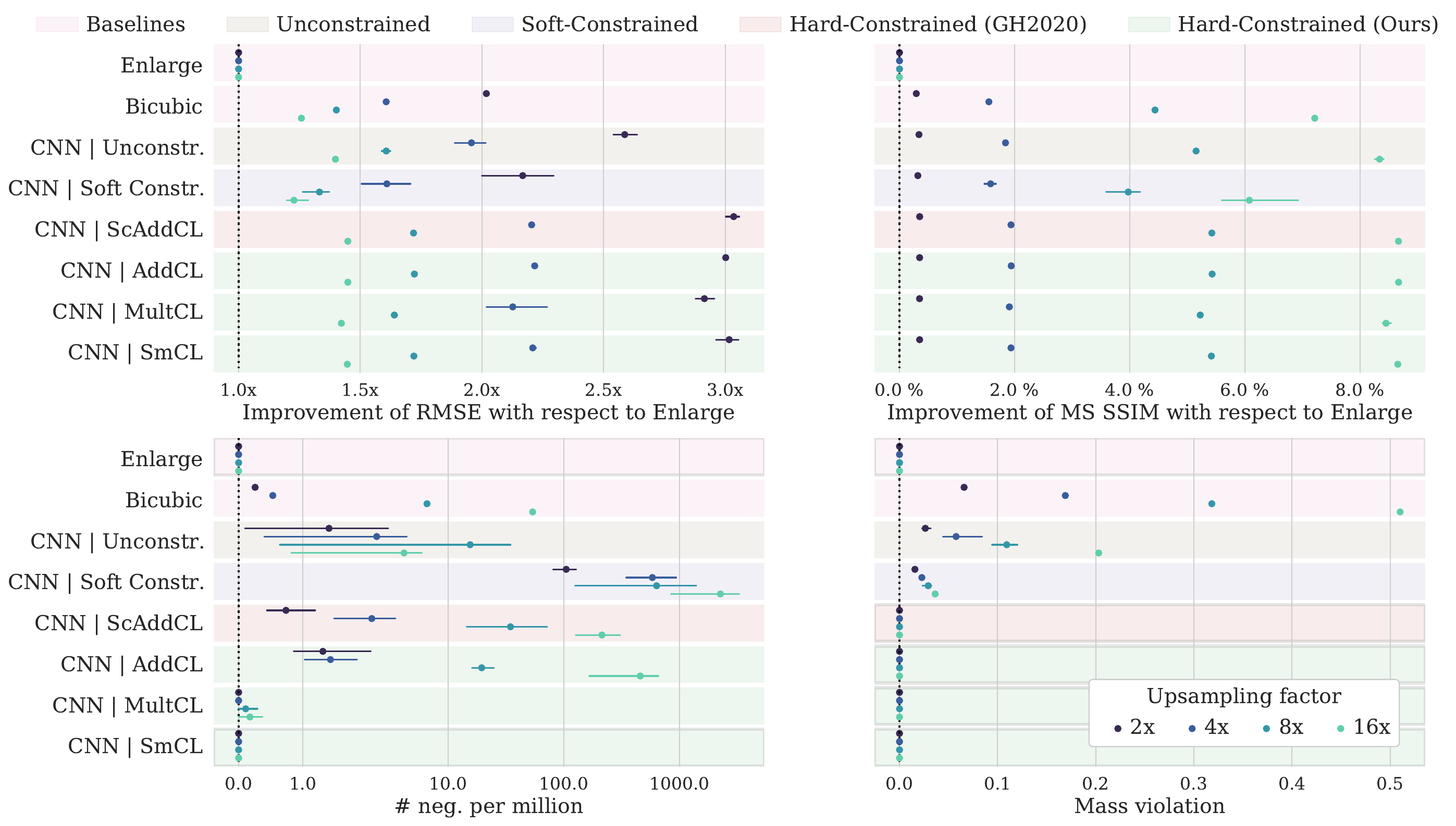}}
\caption{Metrics for different constraining methods applied to an SR CNN, calculated over 10,000 test samples of the water content data. The mean and confidence interval from 3 runs is shown relative to the Enlarge baseline. The framed box indicates a method that achieves zero violation of the physics, no negative pixels or mass conservation up to numerical precision. A table with more metrics can be found in the appendix}
\label{fig:wc}
\end{center}
\end{figure} 

One observation from TFigure \ref{fig:wc} is that the RMSE improvement is better for
lower upsampling factors but the other way around for MS SSIM. A potential explanation: For higher
upsampling factors it gets increasingly difficult to achieve good visual (read
high SSIM) quality, whereas the RMSE is still relatively easy to minimize.
Here, adding the constraint layers have more leverage to improve.

\section*{Appendix D: Additional scores}

We look at additional scores for our water content data set. We investigate the mean bias (mean over the difference for each pixel value of prediction and truth), the peak signal-to-noise ratio (PSNR), the structural similarity index measure, the Pearson correlation (Corr), and the negative mean (the average magnitude of predicted negative values, the average is calculated over all predicted values, including positive, that are set to zero to calculate the negative mean). These metrics show a similar trend then the metrics shown in the main paper: all of them are improved by adding constraints in our architecture. Without or with soft constraining there are small biases appearing in the predictions, but hard constraining removes those biases. PSNR is a function of the MSE and therefore shows the same trend as it. SSIM and correlation give very similar results, with ScAddCL, AddCL, and SmCL showing the best scores. Overall we can see that soft-constraining leads to the most significantly negative predictions, which would cause issues in the context of climate models and predictions.

\begin{table}[htb] 
\caption{More metrics for different constraining methods applied to an SR CNN, calculated over 10,000 test samples. The best scores are highlighted in bold blue, second best in bold.}
\label{cnn_results_2}
\vskip 0.15in
\begin{center}
\begin{small}
\begin{sc}
\begin{tabular}{llllccccc}
\toprule
Data & Fact. & Model & Constraint & Mean bias & PSNR & SSIM &  Corr & Neg mean\\
\midrule
TCW2 &2x & Enlarge & none & \bf{\color{blue}0.000} & 45.36 & 98.65 & 99.75 & \bf{\color{blue}0.000} \\
TCW2 &2x & Bicubic & none & \bf{\color{blue}0.000} & 51.46 & 99.71 & 99.95 & \bf{\color{blue}0.000}\\
TCW2 &2x & CNN & none & $-$0.003 & 53.62 & 99.82 & 99.97 & 0.002\\
TCW2 &2x & CNN & soft & $-$0.002 & 52.07 & 99.74 & 99.94 & 0.192 \\ 
TCW2 & 2x &CNN & AddCL  & \bf{\color{blue}0.000} & 54.91 & \bf{99.85} & \bf{\color{blue}99.98}& 0.002\\
TCW2 & 2x &CNN & ScAddCL & \bf{\color{blue}0.000} & \bf{\color{blue}55.66} &\bf{\color{blue}99.87} & \bf{\color{blue}99.98} & \bf{\color{blue}0.000}\\
TCW2 & 2x &CNN & MultCL & \bf{\color{blue}0.000} & 54.65 & 99.84 & 99.97 & \bf{\color{blue}0.000}\\
TCW2 & 2x &CNN & SmCL & \bf{\color{blue}0.000} & \bf{54.95}  & \bf{99.85} & \bf{99.98} & \bf{\color{blue}0.000} \\
\midrule
TCW4 &4x & Enlarge & none & \bf{\color{blue}0.000} & 39.43 & 94.91 & 98.98 & \bf{\color{blue}0.000}\\
TCW4 & 4x & Bicubic & none & \bf{\color{blue}0.000} & 43.55 & 98.29 & 99.63 & 0.000  \\
TCW4 & 4x & CNN & none & $-$0.015 & 45.26 & 98.70 & 99.74 & 0.001 \\
TCW4 & 4x & CNN & soft & $-$0.001 & 43.55 & 98.15 & 99.59 & 0.546\\ 
TCW4 & 4x &CNN & AddCL & \bf{\color{blue}0.000} & \bf{46.35} & \bf{98.89} &\bf{\color{blue} 99.80}& 0.001\\
TCW4 &4x &CNN & ScAddCL &  \bf{\color{blue}0.000} & \bf{\color{blue}46.42} &  \bf{\color{blue}98.90} & \bf{\color{blue}99.79} & \bf{\color{blue}0.000}\\
TCW4 & 4x &CNN & MultCL &\bf{\color{blue}0.000} & 45.98 & 98.83 & 99.78 & \bf{\color{blue}0.000}\\
TCW4 & 4x &CNN & SmCL & \bf{\color{blue}0.000} &  46.31 &  98.88 & 99.79 & \bf{\color{blue}0.000}\\
 \midrule
TCW8 & 8x & Enlarge & none & \bf{\color{blue}0.000} & 34.84 & 89.08 & 96.95 & \bf{\color{blue}0.000}\\
TCW8 & 8x & Bicubic & none & +0.0001 & 37.77 & 95.40 & 98.50& 0.006  \\
TCW8 & 8x & CNN & none & $-$0.0148 & 38.96& 95.93 & 98.82 & 0.012 \\
TCW8 & 8x & CNN & soft & $-$0.0071 & 37.32 & 94.37 & 98.22 & 0.656 \\ 
TCW8 & 8x &CNN & AddCL & \bf{\color{blue}0.000} & \bf{39.56} & \bf{96.23} & \bf{98.96} & 0.011 \\
TCW8 & 8x &CNN & ScAddCL & \bf{\color{blue}0.000} & \bf{\color{blue}39.58} & \bf{\color{blue} 96.24} & \bf{\color{blue}98.97} & \bf{\color{blue}0.000}\\
TCW8 & 8x &CNN & MultCL & \bf{\color{blue}0.000} & 39.13 & 95.99 & 98.87 & \bf{\color{blue}0.000} \\
TCW8 & 8x &CNN & SmCL & \bf{\color{blue}0.000} &  39.55 & 96.21 & \bf{98.96} &\bf{\color{blue}0.000}\\
  \midrule
TCW16 &16x & Enlarge & none & \bf{\color{blue}0.000} & 30.92 & 85.20 & 92.19 & \bf{\color{blue}0.000} \\
TCW16 & 16x & Bicubic & none & +0.0090 & 32.91 & 91.99 & 95.15 & 0.063\\
TCW16 & 16x & CNN & none & $-$0.0091 & 33.83 & 92.48 & 95.94 & 0.006 \\
TCW16 & 16x & CNN & soft & +0.0115 & 32.70 & 90.45 & 94.63 & 4.233 \\ 
TCW16 & 16x &CNN & AddCL & \bf{\color{blue}0.000} & \bf{\color{blue}34.14} & \bf{92.67} & \bf{\color{blue}96.20} & 0.581 \\
TCW16 &  16x &CNN & ScAddCL & \bf{\color{blue}0.000} & \bf{34.13} & \bf{92.67} & {96.18} & 0.007\\
TCW16 & 16x &CNN & MultCL & \bf{\color{blue}0.000} & 33.98 & 92.54 & 96.07 & \bf{\color{blue}0.000}\\
TCW16 & 16x &CNN & SmCL & \bf{\color{blue}0.000} & \bf{34.13} & \bf{\color{blue}92.68} & \bf{96.19} & \bf{\color{blue}0.000}\\
\bottomrule
\end{tabular}
\end{sc}
\end{small}
\end{center}
\vskip -0.1in
\end{table}

\begin{table}[htb] 
\caption{{Fractional Skill Score (FSS) for different constraining methods and SR CNN applied on the ERA4 water content data, calculated over 10,000 test samples. We look at window sizes 2,4 and 8 and the 95th and 99th percentiles. The best scores are highlighted in bold blue.}}
\label{fss_results}
\vskip 0.15in
\begin{center}
\begin{small}
\begin{sc}
\begin{tabular}{lllcccccc}
\toprule
 Data & Model & Constraint &  & 95perc. &  &  & 99perc. & \\
\midrule
 & & & 2 & 4 & 8 & 2 & 4 & 8\\
 \midrule
TCW4 & Enlarge & none & 0.970 & 0.989 & 0.997 & 0.935 & 0.974 & 0.991 \\
TCW4 & Bicubic & none & 0.971 & 0.987 & 0.994 & 0.935 & 0.969 & 0.986 \\
TCW4 & CNN & none & 0.978 & 0.992 & 0.997 &0.950 & 0.979 & 0.993 \\
TCW4 & CNN & soft & 0.971 & 0.989 & 0.997 & 0.935 & 0.974 & 0.991 \\
TCW4 & CNN& ScAddCL & \bf{\color{blue}0.981} & \bf{\color{blue}0.993} & \bf{\color{blue}0.998} & \bf{\color{blue}0.956} & \bf{\color{blue}0.983} & \bf{\color{blue}0.994} \\
TCW4 & CNN & AddCL  & \bf{\color{blue}0.981} & \bf{\color{blue}0.993} & \bf{\color{blue}0.998} & \bf{\color{blue}0.956} & \bf{\color{blue}0.983} & \bf{\color{blue}0.994} \\
TCW4 & CNN & MultCL & 0.979 & 0.992 & 0.998 & 0.951 & 0.980 & 0.993\\
TCW4 & CNN & SmCL & \bf{\color{blue}0.981} & \bf{\color{blue}0.993} & \bf{\color{blue}0.998} & 0.955 & \bf{\color{blue}0.983} & \bf{\color{blue}0.994}\\
\bottomrule
\end{tabular}
\end{sc}
\end{small}
\end{center}
\vskip -0.1in
\end{table}

\begin{table}[htb] 
\caption{{The variance among super-pixels for different constraining methods and SR CNN applied on the ERA4 water content data, calculated over 10,000 test samples.}}
\label{variance}
\vskip 0.15in
\begin{center}
\begin{small}
\begin{sc}
\begin{tabular}{lllc}
\toprule
 Data & Model & Constraint &  Variance\\
 \midrule
TCW4 & Enlarge & none & 0.00\\
TCW4 & Bicubic & none & 0.85 \\
TCW4 & CNN & none & 1.22 \\
TCW4 & CNN & soft & 0.96\\
TCW4 & CNN& ScAddCL & 1.33 \\
TCW4 & CNN & AddCL  & 1.32 \\
TCW4 & CNN & MultCL & 1.24\\
TCW4 & CNN & SmCL & 1.34 \\
TCW4 & HR & none & 1.65 \\
\bottomrule
\end{tabular}
\end{sc}
\end{small}
\end{center}
\vskip -0.1in
\end{table}


\section*{Appendix E: Additional Visualizations}

\begin{figure*}[htb]
\begin{center}
\vskip 0.1in
\centerline{\includegraphics[width=\columnwidth]{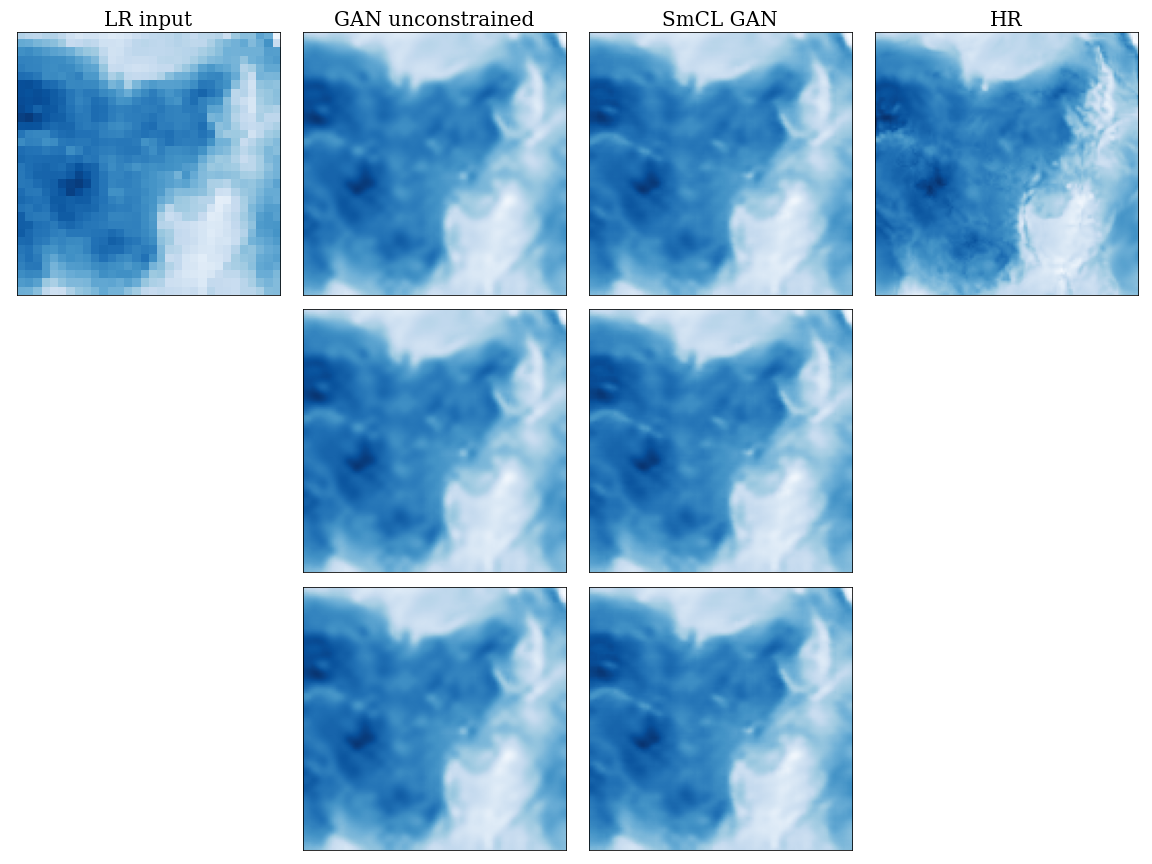}}
\caption{A random sample for the GAN predictions, showing 3 different outputs from the ensemble, constrained and unconstrained. }
\label{gan}
\end{center}
\vskip -0.2in
\end{figure*}

\begin{figure*}[htb]
\begin{center}
\vskip 0.1in
\centerline{\includegraphics[width=\columnwidth]{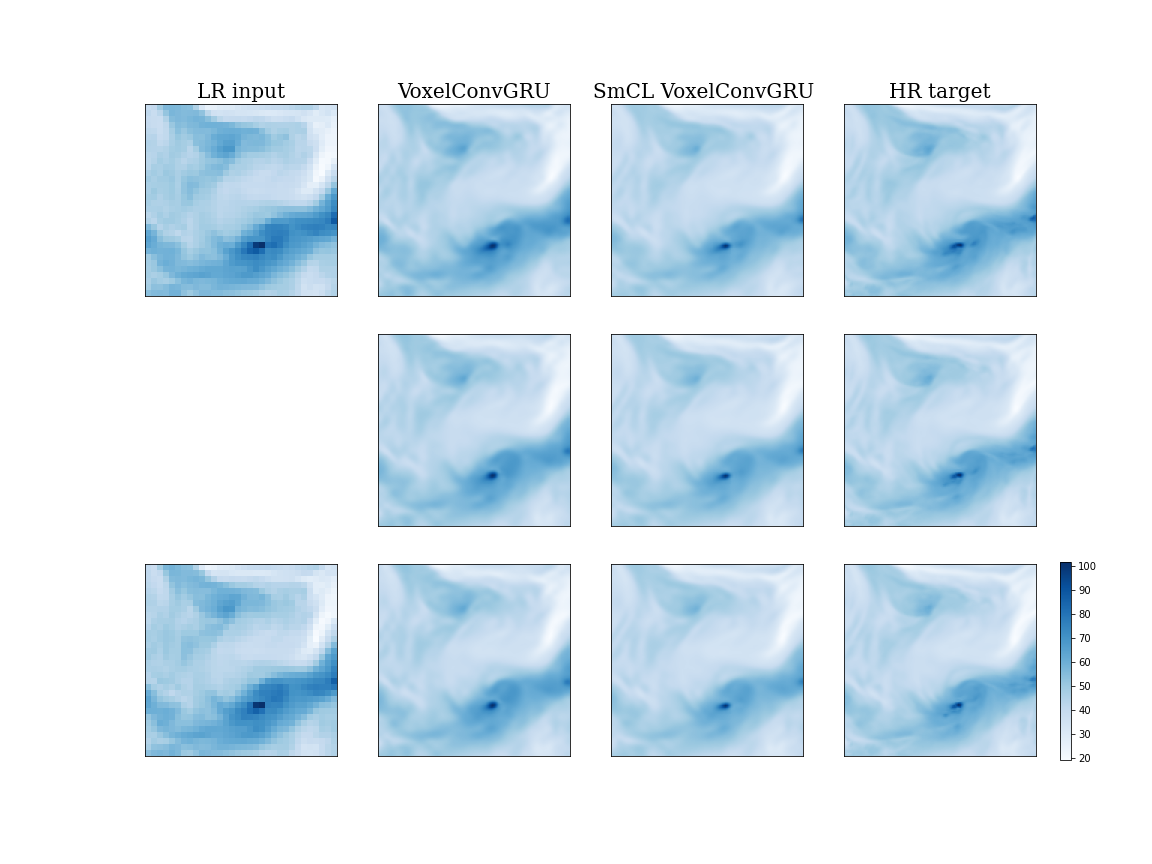}}
\caption{One random test sample and its prediction. Shown here are the two LR input time steps, predictions by both a constrained and unconstrained version of the FlowConvGRU, and the HR sequence as a reference.}
\label{time}
\end{center}
\vskip -0.2in
\end{figure*}

\begin{figure*}[htb]
	\begin{center}
		\vskip 0.1in
		\centerline{\includegraphics[width=\columnwidth]{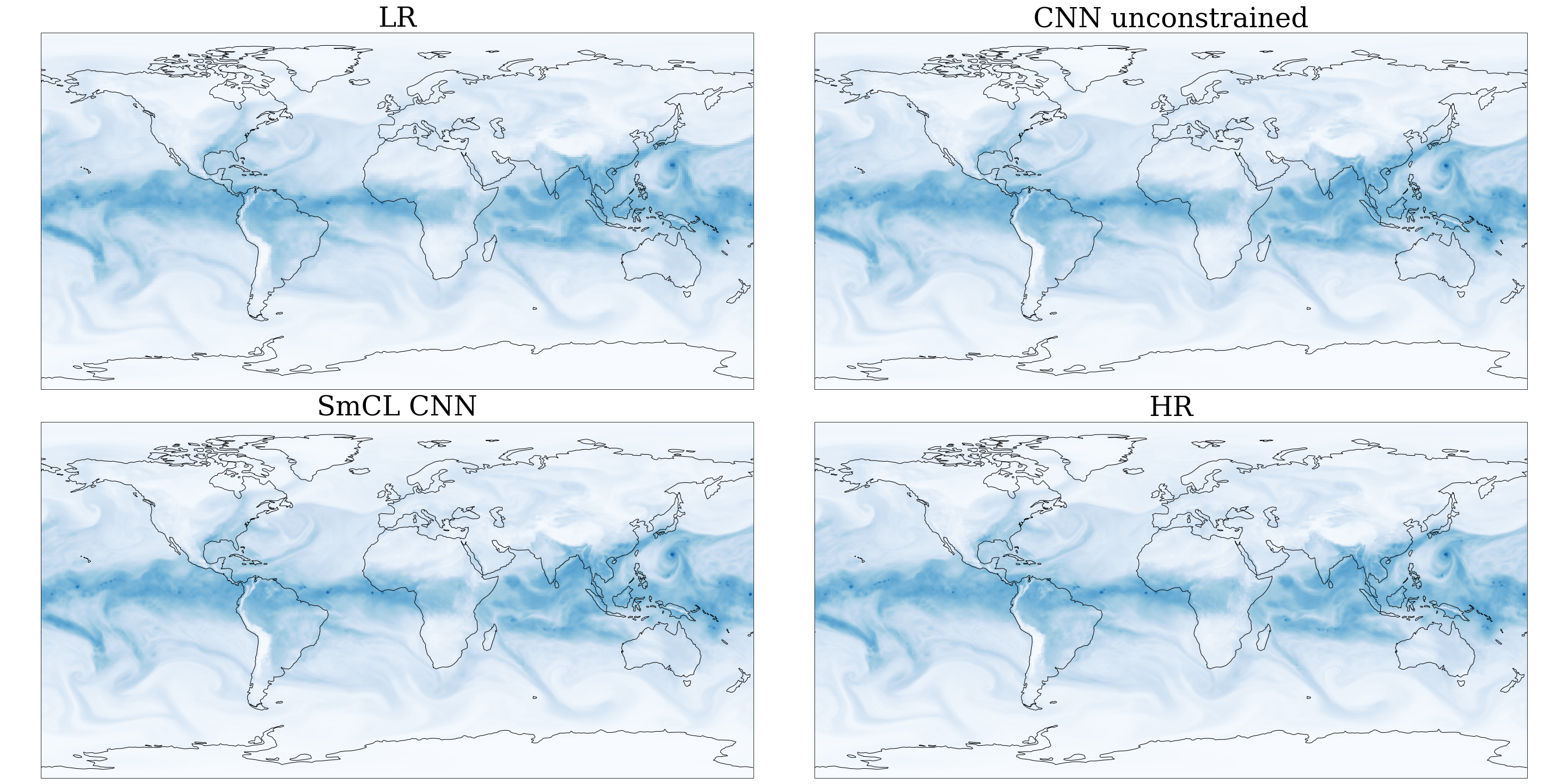}}
		\caption{Global water content data (from data set TCW4): We show LR, unconstrained prediction, softmax-constrained prediction, and HR. The models are applied to one random time step of the test data set, seperately to each 32x32 patch and than combined together to create a global visualization.}
		\label{map}
	\end{center}
	\vskip -0.2in
\end{figure*}

\begin{figure*}[htb]
\begin{center}
\vskip 0.1in
\centerline{\includegraphics[width=\columnwidth]{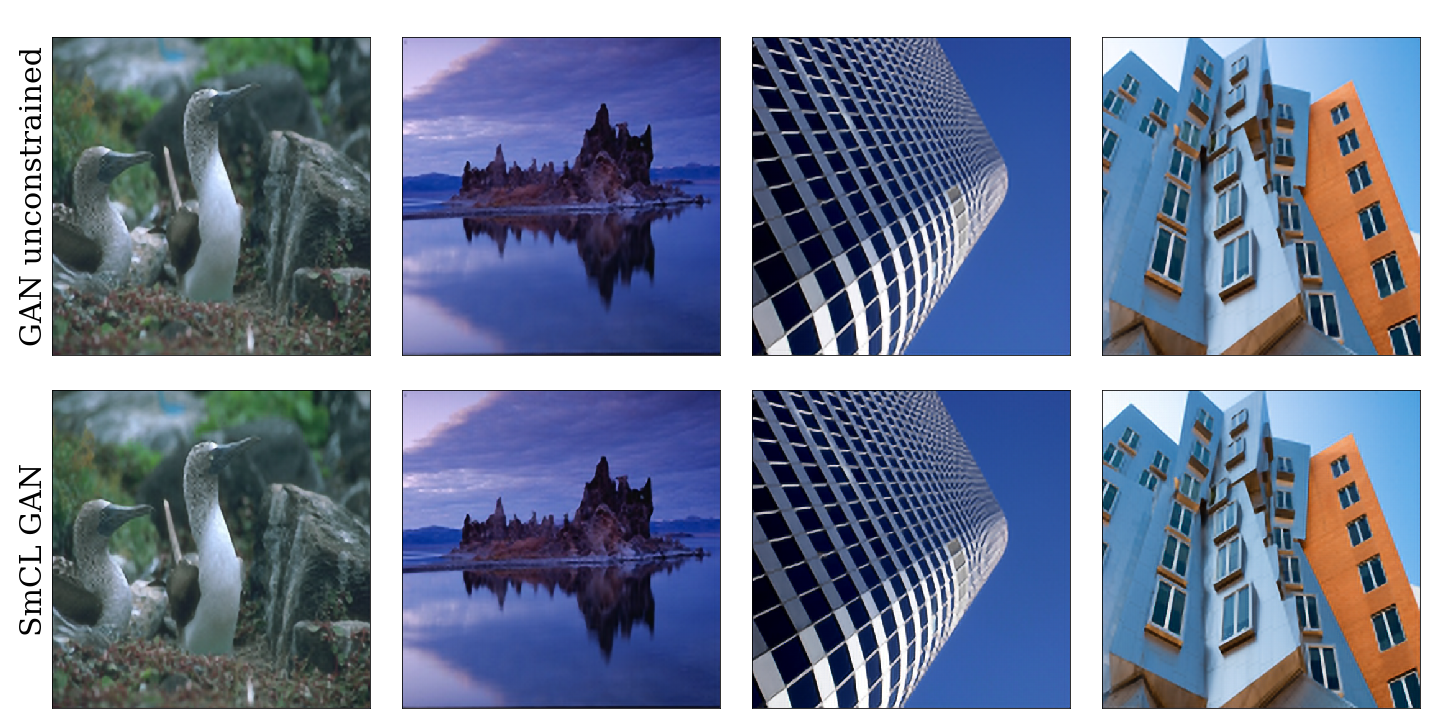}}
\caption{Two random images from both the BSD100 and the Urban100 data sets. The first row shows the unconstrained prediction, the second row the constrained prediction using softmax constraining.}
\label{natty}
\end{center}
\vskip -0.2in
\end{figure*}

Here we present some visualizations, a prediction by the GAN (Figure \ref{gan}), the FlowConvGRU (Figure \ref{time}), unconstrained and constrained example prediction from BSD100 and Urban100 (Figure \ref{natty}), and a global prediction for water content (Figure \ref{map}).

\color{black}

\section*{Appendix F: NorESM data}

\begin{figure*}[htb]
\begin{center}
\vskip 0.1in
\centerline{\includegraphics[width=\columnwidth]{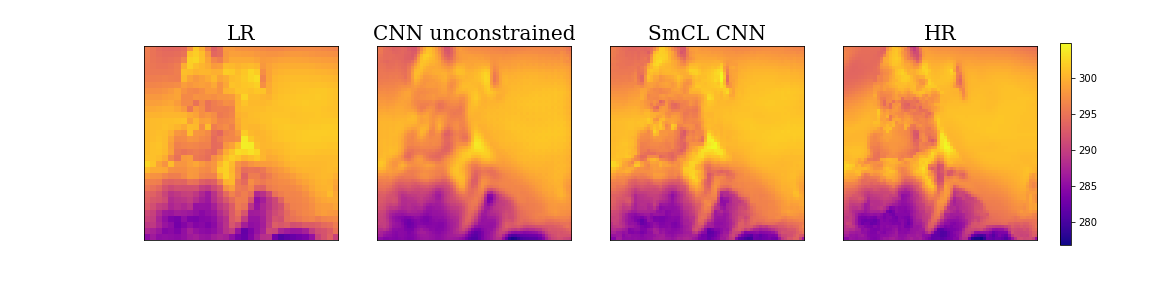}}
\caption{A random sample prediction for the NorESM temperature test data set, we compare an unconstrained CNN and a softmax-constrained CNN here. The constrained prediction looks more similar to the HR ground truth, including more high-frequency features.}
\label{noresm}
\end{center}
\vskip -0.2in
\end{figure*}

\begin{figure*}[htb]
\begin{center}
\vskip 0.1in
\centerline{\includegraphics[width=\columnwidth]{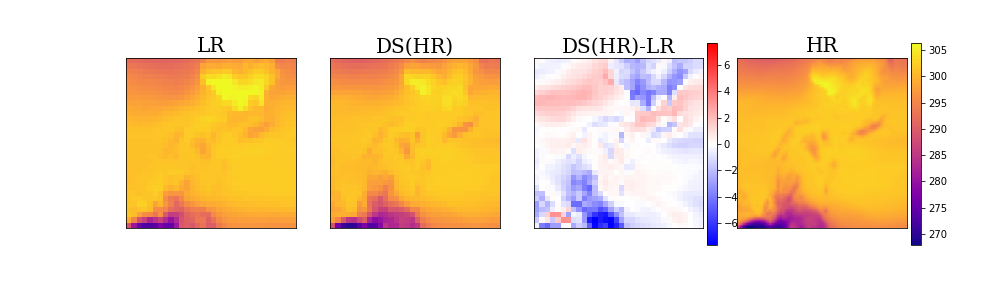}}
\caption{A sample from the NorESM temperature training data set. We compare the low-resolution simulation to the downsampled high-resolution counterpart. It can be observed that the LR and the downsampled HR are significantly different.}
\label{noresm_ds}
\end{center}
\vskip -0.2in
\end{figure*}

Our NorESM data set is based on the second version of the Norwegian Earth System Model (NorESM2), which is a coupled Earth System Model developed by the NorESM Climate modeling Consortium (NCC), based on the Community Earth System Model, CESM2. We build our data set on two different runs: NorESM-MM which has a 1-degree resolution for model components and NorESM2-LM which has a 2-degree resolution for atmosphere and land components. We use the temperature at the surface (tas) and a time period from 2015 to 2100. The scenarios ssp126 and ssp585 are used for training ssp370 for validation and ssp245 for testing. By cropping into $64 \times 64$ and $32\times 32$ pixels, each scenario contains 12k data points.
The results for the NorESM data are shown in Table \ref{noresm_results}: the best scores are in all cases achieved by the unconstrained CNN. This is probably due to the stronger violation of the downscaling constraints between low-resolution and high-resolution samples. We can see a significant difference between the real LR and the HR downsampled, as shown in Figure \ref{noresm_ds}. The violation of the constraints here is 2.48 (RMSE), which is much higher than for the WRF case (0.68). The visual quality of the prediction, on the other hand, seems to be improved by constraining, an example is shown in Figure \ref{noresm}. One potential approach for improvements here could be lat-lon weighted constraining.

\begin{table}[htb] 
\caption{Metrics for different constraining methods applied to the SR CNN, calculated over the test samples of the NorESM data set. The mean is taken over 3 runs. Best scores are highlighted in bold.}
\label{noresm_results}
\vskip 0.15in
\begin{center}
\begin{small}
\begin{sc}
\begin{tabular}{lllcccc}
\toprule
Data & Model & Constraint & RMSE & MAE & MS-SSIM &  Constr. viol. \\
\midrule
Tas NorESM & Enlarge & none & 2.987 & 1.915& 95.96 & \bf{\color{blue}0.000} \\
Tas NorESM &Bicubic & none & 2.910 & 1.864 & 96.36 & 0.073 \\
Tas NorESM &CNN & none & \bf{\color{blue}2.348}& \bf{\color{blue}1.559} & \bf{ \color{blue}96.93} & 1.034 \\
Tas NorESM &CNN& soft & 2.928 & 1.874 & 96.28 & 0.041 \\
Tas NorESM &CNN & AddCL  & 2.885 & 1.847 & 96.45 & \bf{\color{blue}0.000} \\
Tas NorESM &CNN & ScAddCL & \bf{2.884} & \bf{1.846} & \bf{96.46} & \bf{\color{blue}0.000} \\
Tas NorESM &CNN & MultCL & 2.888 & 1.859 & 96.43&  \bf{\color{blue}0.000} \\
Tas NorESM &CNN& SmCL & 2.885 & 1.847 & 96.45 & \bf{\color{blue}0.000} \\
\bottomrule
\end{tabular}
\end{sc}
\end{small}
\end{center}
\vskip -0.1in
\end{table}

\section*{Appendix G: Non-climate data}

\subsection*{Lunar data}

Recent work \citep{lunar} on super-resolution for lunar satellite imagery has shown how deep learning can be used to enhance the captured data to help future missions to the moon. To increase the resolution of images from regions like the south pole, where there is no high-resolution data available, a machine learning-ready data set has been created. It consists of 220,000 images cropped out of the Narrow-Angle Camera (NAC) imagery from NASA's Lunar Reconnaissance Orbiter (LRO); for more details see \citet{lunar}. Here we use a 4x upsampling version of the data set to verify if our constraining methodologies can increase the performance of super-resolution outside of climate science. The average sampling is justified in this case, because the real LR images would be created with summing photon counts in low-light regions.

\subsection*{Natural images}

The standard benchmark data sets for  super-resolution deep learning architectures applied to natural images include the OutdoorScenceTRaining (OST), DIV2K, and Flickr2k data sets for training and Set5, Set14, Urban100, and BSD100 for testing, as for example in \citet{wang2018esrgan}.  Here, we use a version resized to $512\times 512$ pixels for HR and apply average pooling to downsample them. Our constraints depend on the downsample technique used and can not directly be applied to other downsample techniques such as sub-sampling or bicubic interpolation.

\begin{table}[htb] 
\caption{Metrics for different constraining methods applied to the SR-CNN, calculated over the test samples of the lunar data set. The mean is taken over 3 runs. The best scores are highlighted in bold blue.}
\label{lunar_results}
\vskip 0.15in
\begin{center}
\begin{small}
\begin{sc}
\begin{tabular}{lllcccc}
\toprule
Data & Model & Constraint & RMSE & MAE & SSIM  & PSNR\\
\midrule
Lunar &CNN & none & 0.00217& 0.00146& 90.08 & 37.57\\
Lunar &CNN  & SmCL & \bf{\color{blue}0.00213} & \bf{\color{blue}0.00144} & \bf{\color{blue}90.40} & \bf{\color{blue}37.74 }\\
\bottomrule
\end{tabular}
\end{sc}
\end{small}
\end{center}
\vskip -0.1in
\end{table}

\begin{table}[htb] 
\caption{Metrics of the SR-GAN with and without SmCL calculated over the test data sets Set5, Set14, Urban100, BSD100. The better scores are highlighted in bold blue.}
\label{sr_results}
\vskip 0.15in
\begin{center}
\begin{small}
\begin{sc}
\begin{tabular}{lllcccc}
\toprule
Data & Model & Constraint & RMSE & MAE & SSIM & PSNR \\
\midrule
Set5 &SR-GAN & none & 8.57& 4.80&{ 92.48}&29.47 \\
Set5 &SR-GAN  &SmCL & \bf{\color{blue}6.61} & \bf{\color{blue}4.01} & \bf{\color{blue}93.95} & \bf{\color{blue}31.73} \\
\midrule
Set14 &SR-GAN & none & 15.75& 8.82& 86.06&24.28 \\
Set14 &SR-GAN  &SmCL & \bf{\color{blue}14.07} & \bf{\color{blue}8.12} & \bf{\color{blue}87.37} & \bf{\color{blue}25.16} \\
\midrule
Urban100 &SR-GAN & none & 25.00& 14.57& 81.40&20.17 \\
Urban100 &SR-GAN  &SmCL & \bf{\color{blue}23.25} & \bf{\color{blue}13.60} & \bf{\color{blue}83.19} & \bf{\color{blue}20.80} \\
\midrule
BSD100 &SR-GAN  & none & 14.38& 8.28& 85.95&24.97 \\
BSD100 & SR-GAN & SmCL & \bf{\color{blue}13.52} & \bf{\color{blue}7.82} & \bf{\color{blue}87.09} & \bf{\color{blue}25.50} \\
\bottomrule
\end{tabular}
\end{sc}
\end{small}
\end{center}
\vskip -0.1in
\end{table}

\clearpage

\bibliography{sample}

\end{document}